\documentclass[a4paper,11pt]{article}
\pdfoutput=1

\usepackage[noblocks]{authblk}
\usepackage{here}
\usepackage{subfig}
\usepackage[pdftex]{graphicx}
\usepackage{color}
\usepackage{braket}
\usepackage{amssymb, amsmath}
\usepackage{feynmf}
\usepackage{bm}
\usepackage{url}
\usepackage{slashed}
\usepackage{caption}
\usepackage{tabularx}

\usepackage{jcappub}
\usepackage[compat=1.1.0]{tikz-feynhand}
\usepackage{hyperref}
\usepackage[samesize]{cancel}
\usepackage{fancyhdr} 

\newcommand{\lmk}{\left(}
\newcommand{\rmk}{\right)}
\newcommand{\lkk}{\left[}
\newcommand{\rkk}{\right]}
\newcommand{\lnk}{\left\{}
\newcommand{\rnk}{\right\}}

\begin{document}
\title{Revisiting the Chern-Simons interaction during inflation with 
a non-canonical pseudo-scalar}
\author[a,b,c]{Jun'ya Kume}
\author[a,b]{Marco Peloso}
\author[a,b,d]{and Nicola Bartolo}
\affiliation[a]{Dipartimento di Fisica e Astronomia ``G. Galilei'', Universit\`a degli Studi di Padova, via Marzolo 8, I-35131 Padova, Italy}
\affiliation[b]{INFN, Sezione di Padova, via Marzolo 8, I-35131 Padova, Italy}
\affiliation[c]{Research Center for the Early Universe (RESCEU), Graduate School of Science, The University of Tokyo, Hongo 7-3-1
Bunkyo-ku, Tokyo 113-0033, Japan}
\affiliation[d]{INAF, Osservatorio Astronomico di Padova, vicolo dell’ Osservatorio 5, I-35122 Padova, Italy}

\emailAdd{junya.kume@unipd.it}
\emailAdd{marco.peloso@pd.infn.it}
\emailAdd{nicola.bartolo@pd.infn.it}

\subheader{{\rm RESCEU-16/24}}

\abstract{
A Chern-Simons interaction between a pseudo-scalar field and a U(1) gauge field results in the generation of a chiral  gravitational wave background. 
The detection of this signal is contrasted by the fact that this coupling also generates primordial scalar perturbations, on which strong limits exist, particularly at CMB scales. 
In this study, we propose a new extension of this mechanism characterized by a non-canonical kinetic term for the pseudo-scalar. We find that a decrease of the sound speed of the pseudo-scalar field highly suppresses the sourced scalar with respect to the sourced tensor modes, thus effectively allowing for the production of a greater tensor signal. 
Contrary to the case of a canonical axion inflaton, it is in this case possible for the sourced tensor modes to dominate over the vacuum ones without violating the non-Gaussianity constraints from the scalar sector, which results in a nearly totally polarized tensor signal at CMB scales. 
We also study the extension of this mechanisms to the multiple field case, in which the axion is not the inflaton. 
}
\maketitle
\section{Introduction}

Cosmic inflation~\cite{PhysRevD.23.347,Sato:1980yn,Linde:1981mu,PhysRevLett.48.1220,STAROBINSKY198099} has become the standard paradigm for early universe physics in modern cosmology, as it not only resolves classical problems in big bang cosmology but also accounts for the properties of the large-scale structure of the universe. This framework provides a robust mechanism for generating primordial curvature perturbations~\cite{Mukhanov:1981xt,HAWKING1982295,PhysRevLett.49.1110,STAROBINSKY1982175,Abbott:1984fp}, with a remarkable agreement with observations~\cite{Planck:2018jri, Planck:2019kim}. 
Substantial experimental efforts are also devoted to the detection of the tensor perturbations, {\it i.e.}, primordial gravitational waves (GWs), produced during inflation~\cite{POLARBEAR:2017beh, BICEP2:2018kqh, SPT:2019nip, ACT:2020frw, BICEP:2021xfz, SimonsObservatory:2018koc, CMB-S4:2020lpa, Moncelsi:2020ppj, LiteBIRD:2022cnt}. 
There is a guaranteed tensor signal from the accelerated expansion during inflation, due to metric quantum vacuum fluctuations, with a power that is directly proportional to the energy scale of inflation~\cite{Grishchuk:1974ny, Starobinsky:1979ty}. Therefore, its detection would provide us with valuable information on the energy at which inflation took place. This strict relation is no longer true if some (model-dependent) sources produce an additional tensor signal greater than the one from the vacuum fluctuations.
If the inflaton is a pseudo-scalar field, for example, it can couple with massless U(1) gauge fields via a theoretically motivated axial coupling term $\frac{\sigma}{f} F\tilde{F}$. This interaction yields a strong amplification of one polarization mode of the gauge field during inflation, resulting in the production of circularly polarized tensor perturbations~\cite{Anber:2006xt,Sorbo:2011rz}.
Other examples  of this from the literature include the generation of GWs from spectator fields~\cite{Biagetti:2013kwa,Biagetti:2014asa,Fujita:2014oba}, from an effective field theory approach of broken spatial diffeomorphism~\cite{Cannone:2014uqa,Bartolo:2015qvr}, or from particle and string production during inflation~\cite{Cook:2011hg,Senatore:2011sp,Barnaby:2012xt,Carney:2012pk}. 

Such mechanisms for enhancing the tensor signal are actually limited by the fact that these sources also produce scalar perturbations, typically with an amount that exceeds that of the tensor modes. This results in many of these models having a decreased overall tensor-to-scalar ratio and makes it impossible to observe the produced GW once the limits from scalar production, enforced by constraints on non-Gaussianity at Cosmic Microwave Background (CMB) scales~\cite{Planck:2018jri,Planck:2019kim} or on Primordial Black Holes abundance at smaller scales, are taken into account~\cite{Biagetti:2014asa,Barnaby:2012xt,Papageorgiou:2019ecb}.
For instance, in the model where massless U(1) gauge fields are sourced by their axial coupling to an axionic inflaton, strong bounds on the coupling are derived from the scalar production~\cite{Barnaby:2010vf,Barnaby:2011vw,Barnaby:2011qe,Linde:2012bt}, excluding the possibility of observing the sourced tensor signal.
Nevertheless, a decreased ratio of sourced scalar vs. sourced tensor perturbations is produced if the sourcing fields are coupled only gravitationally. 
This feature is present in the model of Refs.~\cite{Barnaby:2012xt,Mukohyama:2014gba}, in which the axion field is a spectator field.~\footnote{Following common terminology, we denote by a {\it spectator field} a field whose background value contributes negligibly to the expansion of the universe and whose fluctuations are not identified with the late-time curvature perturbation.} If the spectator axion is assumed to continuously roll all throughout inflation, strong constraints from scalar generation still apply~\cite{Ferreira:2014zia}. Ref.~\cite{Namba:2015gja} considered instead a spectator axion with a mass of the order of the Hubble rate, that rolls in a typical cosine potential for only a few $e$-folds (typically, $\sim 2 -5$) of inflation. This results in sourced signals localized at the scales that left the horizon while the axion was rolling. This significantly decreases the limits from the CMB non-Gaussianity, so that this mechanism can produce an observable tensor signal at CMB scales for inflation at an arbitrarily low energy scale. This model was recently studied via lattice simulations~\cite{Caravano:2024xsb}.

In this study, we explore an extension of this mechanism, focusing on whether the sourced tensor modes can be observable while satisfying the non-Gaussianity constraints in the scalar sector.
Specifically, we introduce a non-canonical kinetic term for the scalar field and examine how this modification impacts both single-field and multi-field scenarios.
For a standard scalar field (not a pseudo-scalar), extensions of models that consider non-canonical kinetic terms have drawn considerable interest, particularly due to their connections with high-energy theories and significant impact on the scalar perturbations.
For example, in the context of string theory, non-canonical kinetic terms arise naturally in models such as DBI inflation~\cite{Silverstein:2003hf,Alishahiha:2004eh}, where the inflaton dynamics are constrained by an extra-dimensional geometry. Other examples include k-inflation~\cite{Armendariz-Picon:1999hyi,Garriga:1999vw}, G-inflation~\cite{Kobayashi:2010cm}, or ghost condensate~\cite{Arkani-Hamed:2003pdi,Arkani-Hamed:2003juy}.
These models have attracted attention as they can naturally generate detectable non-Gaussian signatures in the scalar sector~\cite{Chen:2006nt}. 

The inclusion of non-standard kinetic term for an axion field was investigated, {\it e.g.}, in the context of inflation~\cite{Maity:2012dx, Maity:2014oza,Domcke:2017fix, Maity:2018ipt,Watanabe:2020ctz,Almeida:2020kaq,Dimastrogiovanni:2023oid,Murata:2024urv} and dark matter production~\cite{Alonso-Alvarez:2017hsz}, leading to predictions that significantly differ from those in standard scenarios. As we will see below, this is also the case in our context. The pseudo-scalar field and its perturbations can attain large inertia due to the non-canonical kinetic term. This allows to suppress the scalar production from the inverse decay of the amplified gauge field while the tensor production remains unaffected. We should note that while models with non-standard axion kinetic terms (in the Einstein frame) have been discussed in the literature, see for instance Refs.~\cite{Domcke:2017fix,Alonso-Alvarez:2017hsz}, we are not aware of any explicit top-down constructions, {\it e.g.} from supergravity or string theory, which are both characterized by a strong impact of the nonstandard kinetic term and under perturbative control.
We leave the investigation of the potential to embed our phenomenological study in a more complete theoretical framework for future work.

The paper is organized as follows. In Sec.~\ref{sec:bg_pert}, we introduce our model which consists of pseudo-scalar field and a U(1) gauge field. 
After describing the inflationary background with non-canonical kinetic term, the amplification of gauge fields is discussed in Sec.~\ref{sec:bkg_gauge} and the full action of scalar perturbations is presented in Sec.~\ref{sec:action_pert}. 
Then, in Sec.~\ref{sec:scalar_sourced}, we discuss observational prospects based on our numerical evaluation of sourced perturbations. We consider the case where the axion is identified with the inflaton in Sec.~\ref{sec:single_field}, and the case where the axion is a spectator field in Sec.~\ref{sec:two_fields}.
Sec.~\ref{sec:discussion} is devoted to discussion. The paper is concluded by three appendices. In App.~\ref{app:scalar_detail} and~\ref{sec:tensor} we present, respectively, details on the computation of the sourced scalar and tensor perturbations. Finally, in App.~\ref{app:comparison} we compare the contributions to the sourced scalar perturbations from the direct Chern-Simons and from the gravitational axion-gauge field couplings.

\section{Gauge field production and action for perturbations}\label{sec:bg_pert}
We consider a system of two spin-zero fields $\phi$ and $\sigma$ with a non-canonical kinetic term. We assume that $\sigma$ is a pseudo-scalar, that is coupled to a U(1) gauge field $A_{\mu}$ via a dimension-5 axial operator.
The lagrangian of the system reads
\begin{equation}
\mathcal{L} = K_{\phi} \left( X \right) + K_{\sigma} \left( Y \right) - V \left( \hat{\phi} ,\, \hat{\sigma} \right) 
- \frac{1}{4} g^{\mu \alpha} g^{\nu \beta} F_{\mu \nu} F_{\alpha \beta} - \frac{1}{8 \sqrt{-g}} \frac{\hat{\sigma}}{f} \epsilon^{\mu \nu \alpha \beta} F_{\mu \nu} F_{\alpha \beta},\label{eq:lag_mat}
\end{equation} 
where $V \left( \hat{\phi} ,\, \hat{\sigma} \right)$ is a generic scalar potential that sustains inflation, $F_{\mu\nu} = \partial_{\mu}\hat{A}_{\nu} - \partial_{\nu}\hat{A}_{\mu}$ is the field strength of the abelian gauge field $A_\mu$, while $\epsilon^{\mu \nu \alpha \beta}$ is the totally antisymmetric Levi-Civita tensor, with $\epsilon^{0123} = 1$. Here, the kinetic terms of the scalar fields, $K_{X/Y}$, are functions of
\begin{equation}
X \equiv - \frac{1}{2} g^{\mu \nu} \partial_\mu \hat{\phi}  \partial_\nu \hat{\phi} \;\;,\;\; 
Y \equiv - \frac{1}{2} g^{\mu \nu} \partial_\mu \hat{\sigma}  \partial_\nu \hat{\sigma},
\end{equation}
and the standard case corresponds to $K_{\phi} = X$ and $K_{\sigma} = Y$. For simplicity, we assume that they depend only on $X$ and $Y$ but not on the field values. Note that our model is complementary to the one studied in Ref.~\cite{Domcke:2017fix}, where the kinetic term takes the form $K_{\sigma} = K(\sigma)Y$ with $K(\sigma)$ being a function only of the field amplitude. 

Throughout this section we consider the presence of two fields, ${\hat \phi}$ and ${\hat \sigma}$, of which only the second one is directly coupled to the gauge field. In Subsection~\ref{sec:single_field} we consider a single-field case, identifying the coupled field ${\hat \sigma}$ with the inflaton, and effectively disregarding the field ${\hat \phi}$. In Subsection~\ref{sec:two_fields} we instead assume that both fields are dynamically relevant, and that ${\hat \phi}$ plays the role of the inflaton, while the coupled field ${\hat \sigma}$ is a spectator field.

To shorten the notation we define a field vector and index
\begin{equation}
\varphi_I = \lmk \phi,\, \sigma \rmk \;\;,\;\; I \in \left\{ \phi ,\, \sigma \right\}.
\end{equation}
The scalar fields are decomposed into the homogeneous part and perturbation as
\begin{equation}
    \hat{\varphi}_I(t,\vec{x}) = \varphi_I(t) + \delta \hat{\varphi}_I(t, \vec{x}).
\end{equation}
We take the mostly positive signature for the Friedmann–Lema\^itre–Robertson–Walker (FLRW) line element, 
\begin{equation}
d s^2 = - d t^2 + a^2 d \vec{x}^2 = a^2 \left[ - d \tau^2 + d \vec{x}^2 \right]. 
\end{equation}
In the following, prime denotes derivative w.r.t. conformal time $\tau$, while dot denotes derivative w.r.t. physical time $t$.

For convenience, we work in the $\hat{A}_0 = \partial_i\hat{A}_i = 0$ gauge and introduce the electromagnetic notation
\begin{equation}
\hat{E}_i \equiv - \frac{1}{a^2} \hat{A}_i' \;\;,\;\; 
\hat{B}_i \equiv \frac{1}{a^2} \epsilon_{ijk} \partial_j \hat{A}_k, 
\end{equation}
although we are not necessarily assuming that the U(1) field is the electromagnetic one. 

As detailed below, the dynamics of the system can be divided into three parts: I) the background consisting of the slowly rolling scalar fields, II) the gauge quanta generated from the background dynamics: $\sigma(t) \to A$, and III) the scalar and tensor perturbations sourced by the gauge field: $A + A \to \delta\sigma, h$. 
Here we describe the inflationary background and the amplification of the gauge field in Sec.~\ref{sec:bkg_gauge}, and then present the full action for the scalar perturbations in Sec.~\ref{sec:action_pert}.

\subsection{Background equations and the vector production}\label{sec:bkg_gauge}

Assuming a negligible backreaction of the amplified gauge fields on the background dynamics (we comment on this assumption at the end of this subsection), the (0,0) and the diagonal (i,j) components of the Einstein equations, and the field equations result in the background equations
\begin{equation}
\begin{aligned}
& \frac{\dot{a}^2}{a^2} = \frac{1}{3 M_p^2} \left[ K_{\phi,1} \dot{\phi}^2 - K_\phi + K_{\sigma,1} \dot{\sigma}^2 - K_\sigma + V \right],\\
& \frac{2 \ddot{a}}{a} + \frac{\dot{a}^2}{a^2} = \frac{1}{M_p^2} \left[ - K_\phi - K_\sigma + V \right], \\
& \left( K_{I,1}+\dot{\varphi}_I^2 \, K_{I,2} \right) \ddot{\varphi_I} + 3 K_{I,1} \frac{\dot{a}}{a} \dot{\varphi_I} + \frac{\partial V}{\partial \varphi_I} = 0,  
\end{aligned}\label{eq:eom_bg}
\end{equation}
where all the quantities related to the scalar fields are evaluated at their background value $\varphi_I(t)$ and we define the $n$-th order derivatives of kinetic terms as
\begin{equation}
K_{\phi,n} \equiv \frac{\partial^n K_\phi}{\partial X^n} \;\;,\;\; 
K_{\sigma,n} \equiv \frac{\partial^n K_\sigma}{\partial Y^n}. 
\end{equation}
As in the standard case, not all the above equations are independent, since they are related by one nontrivial Bianchi identity. 

Despite the presence of non-canonical kinetic terms, one can actually impose slow roll conditions in strong analogy with the standard scenario~\cite{Mukhanov:2005bu}. By combining the first two equations in Eq.~\eqref{eq:eom_bg}, the time derivative of the Hubble parameter $H\equiv \dot{a}/a$ can be expressed as
\begin{equation}
\dot{H} = - \left( \epsilon_\phi + \epsilon_\sigma \right) H^2 \equiv - \epsilon \, H^2 \;, 
\end{equation}
where the slow roll parameters are defined as 
\begin{equation}
\epsilon_I \equiv \frac{K_{I,1} \, \dot{\varphi}_I^2}{2 M_p^2 H^2} \;\;,\;\; 
\epsilon \equiv \epsilon_\phi + \epsilon_\sigma. 
\end{equation}
Therefore, assuming $|K_I / V|, |\epsilon_I| \ll 1$, the system follows quasi-de Sitter expansion as $H \simeq V/3M_p^2\simeq {\rm const.}$
To ensure that such time evolution lasts sufficiently long, one needs to monitor the equations of motion in slow roll. By disregarding the second time derivative, we find
\begin{equation}
\Phi_I \equiv 3 K_{I,1}H\dot{\varphi}_I  + \frac{\partial V}{\partial \varphi_I} \simeq 0. 
\label{slow}
\end{equation}
The time evolution of this quantity can be characterized by the slow-roll parameters as
\begin{eqnarray}
\dot{\Phi}_I + 3 H \Phi_I &=& 3 H^2 M_p \sqrt{2 K_{I,1}} 
\left(
\sum_J \sqrt{\epsilon_J} \, \eta_{JI} - \epsilon \sqrt{\epsilon_I}
\right) \;, 
\end{eqnarray}
where we defined
\begin{equation}
\eta_{IJ} \equiv \frac{1}{3 H^2 \sqrt{K_{I,1}} \sqrt{K_{J,1}}} \, \frac{\partial^2 V}{\partial \varphi_I \partial \varphi_J}. 
\end{equation}
Therefore, in order to assume inflationary background, we require the following slow-roll conditions throughout this work:
\begin{equation}
|K_I/V|,\, \left\vert \epsilon_I \right\vert ,\, \left\vert \eta_{IJ} \right\vert \ll 1.
\end{equation}
Here, we assume these slow-roll conditions to hold, but we do not discuss explicit constructions of slow-roll inflation with nonstandard kinetic terms. We refer the interest reader to Refs.~\cite{Silverstein:2003hf,Alishahiha:2004eh,Armendariz-Picon:1999hyi,Garriga:1999vw,Kobayashi:2010cm,Arkani-Hamed:2003pdi,Arkani-Hamed:2003juy}, for some examples. 

Next, we consider the evolution of the U(1) gauge field sourced by the homogeneous background evolution of the pseudo-scalar field $\sigma(t)$.
Following standard practice, we expand the gauge field operator as 
\begin{equation}
{\hat A}_i(\tau,\,{\vec x})=
\int\frac{d^3k}{\left(2\pi \right)^{3/2}} \, {\rm e}^{i{\vec k\cdot \vec x}}{\hat A}_i(\tau,{\vec k})=
\sum_{\lambda=\pm}\int \frac{d^3k}{\left(2\pi \right)^{3/2}}\left[\epsilon_i^{(\lambda)}(\vec k)\,A_\lambda(\tau,\,\vec k)\,{\hat a}_\lambda \left( \vec k \right) \, {\rm e}^{i{\vec k\cdot \vec x}}+{\mathrm {h.c.}}\right] \;, \label{eq:gauge_operator}
\end{equation}
where the ladder operator ${\hat a}_\lambda \left( \vec k \right)$ satisfies the standard commutation relation
\begin{equation}
\lkk {\hat a}_\lambda \left( \vec k \right),\, {\hat a}_{\lambda^{\prime}} \left( \vec k^{\prime}  \right)\rkk = \delta_{\lambda\lambda^{\prime}}\delta \lmk \vec k - \vec k^{\prime}\rmk  \;, \label{eq:commutation}
\end{equation}
and the circular polarization vectors satisfy 
\begin{equation}
\begin{aligned}
& e_a^{(\lambda)} \left( - {\hat k} \right) = 
e_a^{(-\lambda)} \left( {\hat k} \right) = 
e_a^{(\lambda)*} \left( {\hat k} \right) \;\;,\;\; 
e_a^{(\lambda)} \left( {\hat k} \right)
e_a^{(\lambda')} \left( {\hat k} \right)
= \delta_{\lambda,-\lambda'} \;, \\
& {\hat k} \cdot \vec{e}^{(\lambda)} \left( {\hat k} \right) = 0 \;\;,\;\; 
{\hat k} \times \vec{e}^{(\lambda)} \left( {\hat k} \right) =  -i \lambda \vec{e}^{(\lambda)} \left( {\hat k} \right).
\end{aligned}
\end{equation}
Extremizing the action, one finds the equation of motion of the gauge mode function as
\begin{equation}
\left( \partial_\tau^2 + k^2 + \lambda \frac{2k\xi}{\tau}\right) A_\lambda \left(\tau, \, \vec{k} \right) = 0,\label{eq:eom_gauge_mode}
\end{equation}
to leading order in slow roll (specifically, assuming the de Sitter evolution for the scale factor, $a = - \frac{1}{H \tau}$), where we introduced the parameter
\begin{equation}
    \xi \equiv \frac{\dot{\sigma}}{2Hf} = \frac{1}{2f}\sqrt{2\epsilon_\sigma}\frac{M_{\rm Pl}}{\sqrt{K_{\sigma,1}}},
\label{xi}
\end{equation}
which characterizes the strength of scalar-gauge coupling and is nearly constant during inflation. Notice that $K_{\sigma,1}$ is involved in the second expression.

The equation of motion~\eqref{eq:eom_gauge_mode} admits the tachyonic instability for one of the two polarization modes. For $\xi = {\rm const}.$, the growing mode of the gauge field admits the  analytical solution
\begin{equation}
    A_+(\tau, \, k) = \frac{1}{\sqrt{2k}}e^{\pi \xi/2}W_{-i\xi, 1/2}(2ik\tau),
\end{equation}
where $W_{\alpha, \beta}(z)$ is the Whittaker $W$ function.
In the region $(8\xi)^{-1} \lesssim -k\tau \lesssim 2\xi$ that accounts for the most of the power in the sourced gauge field, the solution can be approximated as~\cite{Anber:2006xt} 
\begin{equation}
\begin{aligned}
    A_+(\tau, \, k) &\cong  \frac{1}{\sqrt{2k}}\lmk \frac{k}{2\xi aH}\rmk^{1/4} e^{\pi \xi - 2\sqrt{2\xi k/(aH)}} \equiv \frac{1}{\sqrt{2k}}\tilde{A}_+(\tau, \, k),\\
    \tilde{A}^{\prime}_+(\tau, \, k) &\cong \sqrt{\frac{2 \xi k}{-\tau} }\tilde{A}_+(\tau, \, k).\label{eq:gauge_mode_approx}
\end{aligned}
\end{equation}
Notice the exponential enhancement $e^{\pi\xi}$, representing a significant amplification of the gauge field for $\xi >  1$. 
As mentioned earlier, the amplified gauge field in turn sources the scalar and tensor perturbations, which inherit features distinguishable from those of vacuum fluctuations.
As commonly done in the literature, we use this approximated mode function~\eqref{eq:gauge_mode_approx} in evaluating the sourced perturbations.

Before concluding this subsection, let us comment on our assumption of negligible backreaction of the produced gauge fields on the background dynamics.
As it is well known (see for instance ~\cite{LopezNacir:2011kk} for a general discussion and ~\cite{Anber:2009ua} for an early study on the axion-gauge system), dissipative effects can have a significant backreaction on the inflationary dynamics. Ref.~\cite{Peloso:2016gqs} worked out the condition under which backreaction can be neglected, obtaining that this is the case for $\xi \lesssim 5$ in the canonical case. We repeated their derivation for the non-canonical case and generalized their conditions to
\begin{equation}
    12.6e^{-\pi\xi} \xi^{3/2} > \lmk c_{s,\sigma}\mathcal{P}_\zeta^{(0)}\rmk^{1/2} \,\, {\rm (\sigma\ is\ inflaton)},
\end{equation}
\begin{equation}
    12.6e^{-\pi\xi} \xi^{3/2} > \lnk (\epsilon_{\phi}/\epsilon_\sigma)c_{s,\sigma}\mathcal{P}_\zeta^{(0)}\rnk^{1/2} \,\, {\rm (\phi\ is\ inflaton)},
\end{equation}
where $c_{s,\sigma}$ is the sound speed of the perturbation of $\sigma$ defined below in Eq.~\eqref{eq:cs_def} and $\mathcal{P}_\zeta^{(0)}$ is the vacuum fluctuation of curvature perturbation.
We see that the value of $c_{s,\sigma} < 1$ provides only the logarithmic correction to the upper limit on $\xi$.
In the canonical case, it has been shown (see for instance ~\cite{Peloso:2022ovc} for an analytic study and references therein for various numerical analyses) that for larger values of $\xi$, backreaction in this model is non-local in time, namely the backreaction term at any given time is significantly affected by the earlier time evolution of the system.
We leave the study of strong backreaction in the non-canonical case for future work.

\subsection{Action of the  scalar perturbations and interactions with the gauge field}\label{sec:action_pert}

We now study scalar perturbations about the background solution discussed in the previous subsection. The action and the resulting computations for the tensor modes are instead presented in Appendix~\ref{sec:tensor}, and they are not affected by the non-canonical kinetic term of the pseudo-scalars. 

Let us start from the quadratic part of the action that is obtained from the first three terms in Eq.~\eqref{eq:lag_mat} and the Einstein-Hilbert term. 
We work in the spatially flat gauge:
\begin{equation}
\delta g_{00} = - a^2 \, 2 \Phi \;\;,\;\; \delta g_{0i} = a^2 \partial_i B \;\;,\;\; \delta g_{ij} = 0 \label{eq:metric_pert}
\end{equation}
and integrate out the non-dynamical modes $\Phi$ and $B$ by applying constraint equations to the full action. 
We introduce the canonical variables and the sound speeds for scalar perturbations
\begin{equation}
v_I \equiv \frac{a \, \sqrt{K_{I,1}}}{c_I} \, \delta \hat{\varphi}_I \;\;,\;\; 
c_{s,I}^2 \equiv \frac{K_{I,1}}{K_{I,1} + \dot{\varphi}_I^2 \,  K_{I,2}}. \label{eq:cs_def} 
\end{equation}
As an illustrative purpose, let us evaluate the expression for the sound speed in some non-canonical theories given in the literature. For example, the power-law kinetic term $K_\phi(X) = X(X/M)^{\alpha-1}$, considered in Refs.~\cite{Unnikrishnan:2012zu,Li:2012vta,Lola:2020lvk}, results in $c_{s,\phi}^2 = (2\alpha - 1)^{-1}$. While we assumed $K_\phi(X)$ to be a function of $X$ (but not $\phi$), the sound speed is similarly defined for a theory with $K_\phi(X,\phi)$. This includes the DBI-inflation model~\cite{Silverstein:2003hf,Alishahiha:2004eh}, where $K_\phi(X,\phi) = f(\phi)^{-1}(1 - \sqrt{1 + 2f(\phi)X})$,
resulting in $c_{s,\phi}^2 = 1 - 2f(\phi)X$.

The resulting action greatly simplifies when we eliminate first derivative of the potential of the scalar fields by enforcing the slow-roll equation~(\ref{slow}). We further expand the action to leading order in slow roll, under the assumption that the $n-$th time derivative of $c_{s,I}$ is of $n-$th order in slow roll. We obtain the following quadratic action for scalar perturbations in momentum space:
\begin{equation}
S_{s2} = \frac{1}{2} \int d \tau d^3 k \left[ v_I^{\prime\dagger} \, v_I + a H \sqrt{\epsilon_I \epsilon_J} \frac{c_{s,I}^2 - c_{s,J}^2}{2 c_{s,I} c_{s,J}} \left( v_I^{\prime\dagger} v_J - v_I^{\dagger} v_J' \right) + a^ 2 {\cal M}_{IJ} v_I^{\dagger} v_J 
\right] \;, 
\label{az2}
\end{equation}
where 
\begin{eqnarray}
{\cal M}_{IJ} &\equiv& \left\{ - c_{s,I}^2 \frac{k^2}{a^2} + H^2 \left[ 2 - \sum_K \epsilon_K - \frac{3 \, \dot{c}_{s,I}}{H \, c_{s,I}} + \frac{3}{2} \left( 1 - c_{s,I}^2 \right)  \sum_K \left( \epsilon_K - \frac{\sqrt{\epsilon_K}}{\sqrt{\epsilon_I}} \, \eta_{IK} \right) \right] \right\} \delta_{IJ} \nonumber\\ 
&& + 3 H^2 \left( \frac{c_{s,I}^2+c_{s,J}^2+2 c_{s,I}^2 c_{s,J}^2}{2 c_{s,I} c_{s,J}} \sqrt{\epsilon_I} \sqrt{\epsilon_J} - c_{s,I} c_{s,J} \eta_{IJ} \right),
\end{eqnarray}
which we find consistent with the expressions in Refs.~\cite{Langlois:2008qf,Fujita:2014oba}.
Also, it is easy to verify that the action agrees with Eq. (3.7) of Ref.~\cite{Namba:2015gja} for the quasi de Sitter evolution $a = \left( \frac{1}{-H\tau} \right)^{\frac{1}{1-\epsilon}}$, for decoupled potential, $\eta_{12} = 0$, and for standard sound speeds, $c_{s,I} = 1$.

Now we consider the terms describing the interactions between the gauge field and scalar perturbations: $A + A\to \delta\sigma$. 
By expanding the last two terms of Eq.~\eqref{eq:lag_mat} up to linear order in the metric perturbations, we find
\begin{eqnarray}
\Delta {\cal L}_{sAA} &=& -  \left[ \frac{\Phi}{2} \left(  \hat{E}_i \hat{E}_i + \hat{B}_i \hat{B}_i \right) - \epsilon_{ijk} \hat{E}_i \hat{B}_j \partial_k B  - \frac{\delta \hat{\sigma}}{f} \hat{E}_i \hat{B}_i \right] \nonumber\\
&=& -Q_1 \, \Phi + Q_{2,k} \, \partial_k B + Q_{3} \,  \frac{\delta\hat{\sigma}}{f} \;, 
\end{eqnarray}
where we have introduced the three composite fields
\begin{equation}
\begin{aligned}
Q_1\left( x \right) & \equiv  \frac{1}{2} \left( \hat{E}_i \, \hat{E}_i + \hat{B}_i \, \hat{B}_i \right) \;, \\
Q_{2,k}\left( x \right) & \equiv  \epsilon_{ijk} \hat{E}_i \hat{B}_j \;, \\
Q_{3}\left( x \right) & \equiv  \hat{E}_i \hat{B}_i.
\end{aligned}\label{eq:source_Q}
\end{equation}
We also include the scalar metric contributions from the Einstein-Hilbert term, and we integrate out the non-dynamical metric perturbations $\Phi$ and $B$. We then find, to leading order in slow-roll, the contribution to the momentum space action 
\begin{equation}
\Delta {\cal S}_{sAA} = 
\frac{1}{2} \int d \tau d^3 k \lkk
-\frac{\sqrt{K_{I,1}} \dot{\varphi}_I}{2 M_p^2 H} \frac{1}{ac_{s,I}}
\lmk c_{s,I}^2 a^4 Q_1 - \frac{i k_k}{k^2}(a^4Q_{2,k})' \rmk 
v_I^{\dagger}
+ \frac{c_{s,\sigma}}{\sqrt{K_{\sigma,1}}} a^3\frac{Q_{3}}{f} v_{\sigma}^{\dagger} 
+{\rm h.c.}
\rkk,
\label{az-mix}
\end{equation}
where h.c. denotes the Hermitian conjugate of the preceding terms.
In the standard scenario where $c_{s,\sigma} = K_{\sigma,1} = 1$, the inverse decay via the direct Chern-Simons interaction, proportional to $Q_3$, dominates over the contribution from gravitational couplings, proportional to $Q_1$ and $Q_2$.
In fact, the scalar perturbation produced in this way is highly non-Gaussian and could be in tension with the null detection of primordial non-Gaussianity in the curvature perturbation. 
As shown in Refs~\cite{Barnaby:2010vf,Barnaby:2011vw,Barnaby:2011qe,Linde:2012bt,Ferreira:2014zia, Planck:2015zfm, Planck:2019kim}, this puts a strong upper bound on $\xi$, prohibiting the sourced component (both for scalar and tensor modes) to dominate over the vacuum fluctuation.

In the present case, however, the strength of the $Q_3$ coupling is directly proportional to $c_{s,\sigma}$, and therefore this term is suppressed at small sound speed. The effect of this suppression will be manifest below, once we solve the equation of motion in terms of the curvature perturbation $\zeta$. Since the sourced tensor perturbation is not affected by the scalar sound speed, a small sound speed results in an increased tensor-to-scalar ratio of the sourced perturbations. 

\section{Consequence of the non-canonical kinetic term}\label{sec:scalar_sourced}

In this section we study the effects of the non-canonical kinetic term of the scalar field(s) on the sourced curvature perturbations.  
Our main result is that the these sourced perturbations are suppressed at small sound speed. Although our explicit computations are performed for constant $\xi$, we present a general argument that is valid also if $\xi$ exhibits a time variation, as for instance in the model of Ref.~\cite{Namba:2015gja}.
We first consider the case where the axion is identified with the inflaton in Sec.~\ref{sec:single_field}.
We evaluate the sourced curvature perturbation and discuss the observational constraints. 
Then, in Sec.~\ref{sec:two_fields}, we consider the two-field case where the axion is a spectator field.

Before proceeding, let us summarize the observables of our interest and the characteristics of vacuum fluctuations.
Assuming that the field $\varphi_I$ dominates the final energy density and perturbations, $v_I$ is related to the curvature perturbation $\zeta$ as
\begin{equation}
    \hat{\zeta}\lmk\tau,\vec{k}\rmk \simeq -\frac{H}{\dot{\varphi_I}}\delta\hat{\varphi}_I\lmk\tau,\vec{k}\rmk = \frac{c_{s,I} H \tau}{\sqrt{2 \epsilon_{I}} M_p} \, v_{I}.\label{eq:phi_to_zeta}
\end{equation}
Then we define the power spectrum and bi-spectrum of the curvature perturbation as 
\begin{align}
\frac{k^3}{2 \pi^2} \left\langle \hat{\zeta} \left( 0^- ,\, \vec{k} \right) \hat{\zeta} \left( 0^- ,\, \vec{k}' \right) \right\rangle \equiv 
\delta^{(3)}\left( \vec{k}+\vec{k}' \right)
P_{\zeta} \left( k \right), \label{eq:def_power}
\end{align}
\begin{equation}
\left\langle 
\hat{\zeta} \left( 0^- ,\, \vec{k}_1 \right)
\hat{\zeta} \left( 0^- ,\, \vec{k}_2 \right)
\hat{\zeta} \left( 0^- ,\, \vec{k}_3 \right)
\right\rangle
\equiv \delta^{(3)} \left( \vec{k}_1 + \vec{k}_2 + \vec{k}_3 \right) 
F\lmk \vec{k}_1,\, \vec{k}_2,\, \vec{k}_3\rmk. \label{eq:def_bispec}
\end{equation}
In particular, the bi-spectrum sourced by the amplified gauge fields has a nearly equilateral shape~\cite{Barnaby:2011vw,Barnaby:2011qe,Barnaby:2012xt}, and, on an exact equilateral configuration, it gives the nonlinear parameter~\cite{Barnaby:2011vw} 
\begin{equation}
    f_{\rm NL}^{\rm eff} = \frac{10}{9(2\pi)^{5/2}}\frac{k^6}{P_{\zeta}^2}F\vert_{|\vec{k}_1| = |\vec{k}_2| = |\vec{k}_3| = k} \;. \label{eq:fNL_def}
\end{equation}

For standard kinetic term, the sourced non-Gaussianity is enhanced in the equilateral configuration, since the sourcing gauge-fields are mostly enhanced at slightly super-horizon scales, and then they redshift away. Therefore, at any given moment during inflation only gauge modes of size slightly greater than the horizon are present, sourcing correlations between scalar modes of comparable wavelength. As we show at a technical level in Appendix~~\ref{app:scalar_detail} (in particular, see the discussion below Eq.~\eqref{Tzeta-ana}) these features do not change in presence of non-standard kinetic terms. As a consequence, a small sound speed of the axion field rescales (and, particularly, suppresses) the amplitude of the bi-spectrum, but it does not significantly modify its shape. 

We also evaluate the tensor-to-scalar ratio
\begin{equation}
    r = \frac{P_+ + P_-}{ P_{\zeta}},
\end{equation}
where the power spectrum $P_{\pm}$ of each tensor polarization is defined analogously to the curvature perturbation (see App.~\ref{sec:tensor}).

In our system, the perturbations are the sum of two statistically uncorrelated contributions.
Namely, the contributions sourced by the gauge field adds up to the vacuum fluctuation, which has the mode function 
\begin{equation}
    v_{\rm BD}(\tau,\,k) = \frac{1}{\sqrt{2 c_{s,I} k}} \left( 1 - \frac{i}{c_{s,I} k \tau} \right) {\rm e}^{-i c_{s,I} k \tau},\label{eq:BD}
\end{equation}
that generalizes the Bunch-Davies vacuum to the case of nonstandard sound speed. This results in the vacuum power spectrum at super-horizon scales
\begin{equation}
{\cal P}_{\zeta}^{(0)} = \frac{k^3}{2 \pi^2}
\frac{c^2_{s,I} H^2 \tau^2}{2 \epsilon_{I} M^2_p}
\left\vert v_{\rm BD} \left( \tau = 0^- \right) \right\vert^2 = \frac{H^2}{8 \pi^2 c_{s,I} \epsilon_{I} M_p^2},
\label{Pz0}
\end{equation}
and in the vacuum tensor-to-scalar ratio
\begin{equation}
 r_{\rm vac} = \frac{{\cal P}_+^{(0)} + {\cal P}_-^{(0)}}{{\cal P}_\zeta^{(0)}} = 16c_{s,I} \epsilon_{I}.
\end{equation}

We also note that, in contrast to the canonical field case, the vacuum contribution can have sizable non-Gaussianity for smaller $c_{s,I}$, as lowering this parameter enhances the strength of the self-interactions.
For a general single field model, for example, the upper bound on $f_{\rm NL}^{\rm equil}$ yields the lower bound $c_{s,I} > 0.021$ on the sound speed at 95\% confidence level~\cite{Planck:2019kim}, and possibly a stronger bound once loop corrections are also taken into account~\cite{Kristiano:2021urj,Kristiano:2022zpn}.

\subsection{Pseudo-scalar inflaton with non-canonical kinetic term}\label{sec:single_field}

Here we assume that the field $\sigma$ is the inflaton field, while $\phi$ plays no role. In practice, we remove all terms with index $\phi$ from the above formal expressions. For computational convenience, we redefine the source terms in Eq.~\eqref{eq:source_Q} as 
\begin{equation}
\begin{aligned}
J_{1,2}\left( \tau ,\, \vec{k}\right) \equiv &c_{s,\sigma}^2 \, \frac{a^4 \left( \tau \right) \, Q_1 \left( \tau ,\, \vec{k} \right)}{k^4} - \frac{\partial}{\partial \tau} \left[ \frac{a^4 \left( \tau \right) \, \frac{i k_i}{k^2} Q_{2,i} \left( \tau ,\, \vec{k} \right)}{k^4} \right],\\
J_3\left( \tau ,\, \vec{k} \right) \equiv& \frac{a^4 \left( \tau \right) Q_3 \left( \tau ,\, \vec{k} \right)}{k^4}, 
\end{aligned}\label{eq:def_J}
\end{equation}
to find the equation of motion of the canonical field as 
\begin{equation}
v_\sigma''+\left( c_{s,\sigma}^2 k^2-\frac{2}{\tau^2} \right) v_\sigma = - \frac{1}{a c_{s,\sigma}} 
\lmk
\sqrt{\frac{\epsilon_\sigma}{2}} \, \frac{1}{M_p} 
k^4 J_{1,2} - \frac{c_{s,\sigma}^2}{\sqrt{K_{\sigma,1}}f}k^4J_3
\rmk, 
\label{eq:single_pert_eom}
\end{equation}
where we only keep the leading terms in a slow-roll expansion. The resulting mode can be decomposed into the homogeneous solution $v_{\sigma}^{(0)}$ (corresponding to the vacuum mode function~\eqref{eq:BD}) and the particular solution $v_{\sigma}^{(1)}$ of Eq.~\eqref{eq:single_pert_eom}.
To obtain the latter, we introduce the Green function 
\begin{align}
G_{c_s k} \left( \tau ,\, \tau' \right) =& \, 
{\tilde G}_{c_s k} \left( \tau ,\, \tau' \right) \theta \left( \tau - \tau' \right) \;, \nonumber\\
{\tilde G}_{c_s k} \left( \tau ,\, \tau' \right) =&\,  
\frac{\pi}{2} \sqrt{\tau \tau'} \left[ J_{3/2} \left( - c_s k \tau \right) Y_{3/2} \left( - c_s k \tau' \right) - Y_{3/2} \left( - c_s k \tau \right) J_{3/2} \left( - c_s k \tau' \right) \right],
\end{align}
which satisfies
\begin{equation}
\left( \frac{\partial^2}{\partial\tau^2}  + k^2c_s^2 - \frac{2}{\tau^2} \right) G_{c_s k} \left( \tau ,\, \tau' \right) = \delta \left( \tau - \tau' \right).
\end{equation}
As we are interested in the super-horizon curvature perturbation, we can take the limit
\begin{equation}
\tilde{G}_{c_s k} \left( 0^- ,\, \tau' \right) = \frac{c_s k \tau' \cos \left( c_s k \tau' \right) - \sin \left( c_s k \tau' \right)}{c_s^3 k^3 \tau \tau'}, 
\end{equation}
which leads to the formal solution of the sourced component of $\zeta$ 
\begin{align}
\hat{\zeta}^{(1)} \left( 0^- ,\, \vec{k} \right) 
&=\frac{H^2}{M_p^2} \, \int_{-\infty}^{0^-} d \tau' \, k \; \frac{c_{s,\sigma} k \tau' \cos \left( c_{s,\sigma} k \tau' \right) - \sin \left( c_{s,\sigma} k \tau' \right)}{c_{s,\sigma}^3} 
\left[ \frac{1}{2} J_{1,2} \left( \tau' ,\, \vec{k} \right)  - \frac{c_{s,\sigma}^2 \,  \xi}{\epsilon_\sigma} J_3 \left( \tau' ,\, \vec{k} \right) \right] \label{eq:zeta_single_cs_1}\\
&\simeq
\frac{H^2}{M_p^2} \, \int_{-\infty}^{0^-} d \tau' \, k \;
\lmk - \frac{k^3\tau^{\prime 3}}{3}\rmk 
\left[ \frac{1}{2} J_{1,2} \left( \tau' ,\, \vec{k} \right)  - \frac{c_{s,\sigma}^2 \,  \xi}{\epsilon_\sigma} J_3 \left( \tau' ,\, \vec{k} \right) \right]. 
\label{eq:zeta_single_cs_1_app}
\end{align}
The approximation in the second line works for $\xi \gtrsim {\cal O}(1)$, since in this case the $\tau^{\prime}$-integral in Eq.~\eqref{eq:zeta_single_cs_1} has most of its support in the region $|k\tau^{\prime}| \ll 1$. We note that taking a sound speed of the scalar modes smaller than unity does not affect the formal expressions of the gauge mode solutions, although the nonstandard kinetic term affects the evolution of $\sigma$ and so of the coefficient $\xi$ given by Eq.~(\ref{xi}). This influences the source of both scalar and tensor perturbations, and so it does not by itself impact significantly the tensor-to-scalar ratio $r$. The main effect of the smaller-than-unity sound speed on the value of $r$ is due to the suppression of the coupling strength of the direct interaction, as manifest in Eq.~\eqref{eq:zeta_single_cs_1_app}.
One can intuitively understand that this suppression is due to the large inertia of the scalar perturbation, which arises from the non-canonical kinetic term and makes the excitation of scalar mode less efficient than in the standard case. 

Now we discuss the observational constraints on the sourced perturbations, assuming $\xi = const.$ during the inflation.
Following Ref.~\cite{Barnaby:2011vw}, we express the sourced curvature power spectrum as
\begin{equation}
\begin{aligned}
    {\cal P}_\zeta^{(1)} \left( k \right) \equiv \left[ {\cal P}_\zeta^{(0)} \right]^2 e^{4\pi\xi} f_{2,\zeta} \left[\epsilon_\sigma,\,  c_{s,\sigma},\, \xi\right],
\end{aligned}\label{eq:source_const_xi}
\end{equation}
and the bi-spectrum evaluated on an exactly equilateral configuration as 
\begin{equation}
F\vert_{|\vec{k}_1| \equiv  |\vec{k}_2| = |\vec{k}_3| = k} = \frac{\lkk {\cal P}_{\zeta}^{(0)}\rkk^3}{k^6}e^{6\pi\xi}f_{3,\zeta}\left[\epsilon_\sigma,\,  c_{s,\sigma},\, \xi\right],
\label{F-f3id}
\end{equation}
where $f_{2,\zeta}$ and $f_{3,\zeta}$ are numerically evaluated as Eqs.~\eqref{eq:f2_full} and~\eqref{eq:f3_full}.
Using these results and Eq.~\eqref{eq:fNL_def}, we identify the region of parameter space $\{\epsilon_\sigma, \, c_{s,\sigma}, \, \xi\}$ where the observational bound on scalar non-Gaussianity is satisfied. 
We note that within this region, Eq.~\eqref{eq:f2_full} implies that the sourced contribution to the scalar power spectrum is subdominant. For example, we find ${\cal P}_\zeta^{(1)}/{\cal P}_\zeta^{(0)} \lesssim 0.04$ when the non-Gaussianity bound is saturated, highlighting the strongly non-Gaussian nature of the sourced component.

For the standard equilateral template for a single field inflation, Planck obtained the $-120 \leq f_{\rm NL}^{\rm equil} \leq 68$ interval at 95\% confidence level~\cite{Planck:2019kim}.
As discussed in Ref.~\cite{Barnaby:2011vw}, the difference between the equilateral template used in that analysis and the shape of the bi-spectrum in our model can be accounted for by increasing the experimental error by the inverse of the ``cosine factor'' between the two shapes~\cite{Babich:2004gb}, which in the case of equilateral vs. inverse decay non-Gaussianity amounts to 0.93~\cite{Barnaby:2011vw}. Therefore, the bound recasted as $-129 \leq f_{\rm NL}^{\rm eff} \leq 73$ can be imposed on the expression (\ref{eq:fNL_def}) to find the maximum allowed value of $\xi$, which we denote $\xi_{\rm max}$. For the standard scenario with a canonical kinetic term, one can actually find dedicated analyses in Refs.~\cite{Planck:2015zfm,Planck:2019kim} where the shape of non-Gaussianity from inverse decay were taken into account. Instead of performing such a dedicated analysis, here we simply adopt the approach in Ref.~\cite{Barnaby:2011vw} by rescaling the limit on equilateral non-Gaussianity through the cosine factor.

\begin{figure}[ht!]
\centerline{
\includegraphics[width=0.7\textwidth,angle=0]{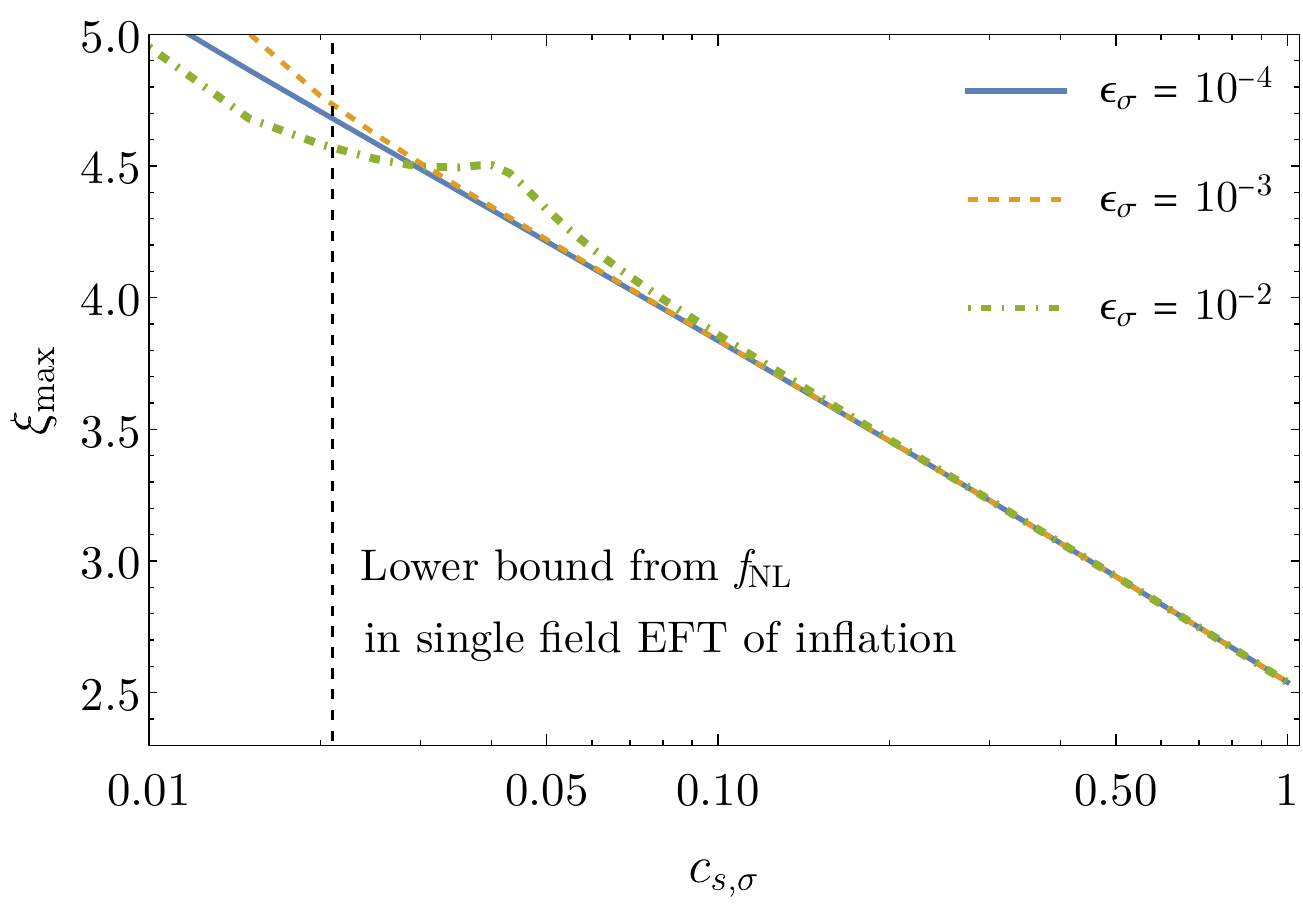}
}
\caption{Maximum allowed value for $\xi$ as a function of $c_s$ compatible with the non-Gaussianity limits. We consider three different values of $\epsilon_{\sigma}$. Deviation from a straight line in the figure are due to the gravitational contributions  ($J_{1,2}$) that interfere more significantly with the direct one ($J_3$) at increasing $\epsilon_{\sigma}$.
}
\label{fig:ximax-cs}
\end{figure}

In Fig.~\ref{fig:ximax-cs}, we plot the values of $\xi_{\rm max}$ as a function of $c_{s,\sigma}$, for different values of $\epsilon_{\sigma}$.
In deriving these limits, the total power spectrum $P_\zeta = {\cal P}_\zeta^{(0)} + {\cal P}_\zeta^{(1)}$ has been fixed to the measured value $2.1\times10^{-9}$~\cite{Planck:2018jri}.
Fig.~\ref{fig:ximax-cs} exhibits a nearly straight line $\xi_{\rm max}-c_{s,\sigma}$ dependence at large $c_{s,\sigma}$. This can be understood by the dominance of the direct source $J_3$ (in Eq.~\eqref{eq:zeta_single_cs_1_app}) in this regime. When this source dominates, only the first term in Eq.~(\ref{eq:f3_full}) is relevant. When this is inserted in Eq.~(\ref{F-f3id}), and then in Eq.~(\ref{eq:fNL_def}), this leads to the scaling ${\rm e}^{6 \pi \xi_{\rm max}} \, c_{s,\sigma}^9/\xi_{\rm max}^9$, which is well respected by result shown in the high $c_{s,\sigma}$ part of the figure. On the other hand, at relatively large $\epsilon_{\sigma}$  and relatively small $c_{s,\sigma}$, the gravitational $J_{1,2}$ source in Eq.~\eqref{eq:zeta_single_cs_1_app} is no longer negligible with respect to the direct source $J_3$, leading to the departure from the nearly straight line that is visible in the figure in this regime. 

As a reference, we also show in Fig.~\ref{fig:ximax-cs} the lower bound on $c_{s,\sigma} > 0.021$ derived from the Planck $f_{\rm NL}$ bound applied on the single field effective field theory of inflation. Comparing the resulting $\xi_{\rm max}$ at this value of the sound speed and at $c_{s,\sigma} = 1$, we see that this is possible to increase the upper limit of $\xi_{\rm max}$ by nearly a factor of two, while still respecting the limit from non-Gaussianity.

In fact, a factor of two yields a drastic change in the observability of the tensor-to-scalar ratio, as this quantity depends on $\xi$ exponentially in the regime in which the sourced modes dominate the tensor contribution, while they remain subdominant in the scalar sector. While this cannot occur for $c_{s,\sigma} = 1$, we now show that this can happen at sufficiently small sound speed. Similarly to the scalar perturbation, we express the sourced tensor power spectrum as
\begin{equation}
\begin{aligned}
    {\cal P}_+^{(1)} \left( k \right)
    \simeq
    \epsilon_{\sigma}^2c_{s,\sigma}^2
    \left[ {\cal P}_\zeta^{(0)} \right]^2 e^{4\pi\xi} f_{h,+} \left[\epsilon_\sigma,\,  c_{s,\sigma},\, \xi\right],
\end{aligned}\label{eq:source_tensor}
\end{equation}
where we only include the dominant $+$ mode, and where the fitting function of $f_{h,+}$ is given in Eq.~\eqref{eq:fh_fit}. Using these relations, the total tensor-to scalar-ratio evaluates to
\begin{equation}
r \simeq \frac{{\cal P}_+^{(0)} + {\cal P}_-^{(0)} + {\cal P}_+^{(1)}}{P_{\zeta}} =
r_{\rm vac}\frac{{\cal P}_{\zeta}^{(0)}}{P_{\zeta}} + \epsilon_{\sigma}^2c_{s,\sigma}^2\frac{\left[ {\cal P}_\zeta^{(0)} \right]^2}{P_{\zeta}}e^{4\pi\xi} f_{h,+}.
\end{equation}

\begin{figure}[ht!]
\centerline{
\includegraphics[width=0.37\textwidth,angle=0]{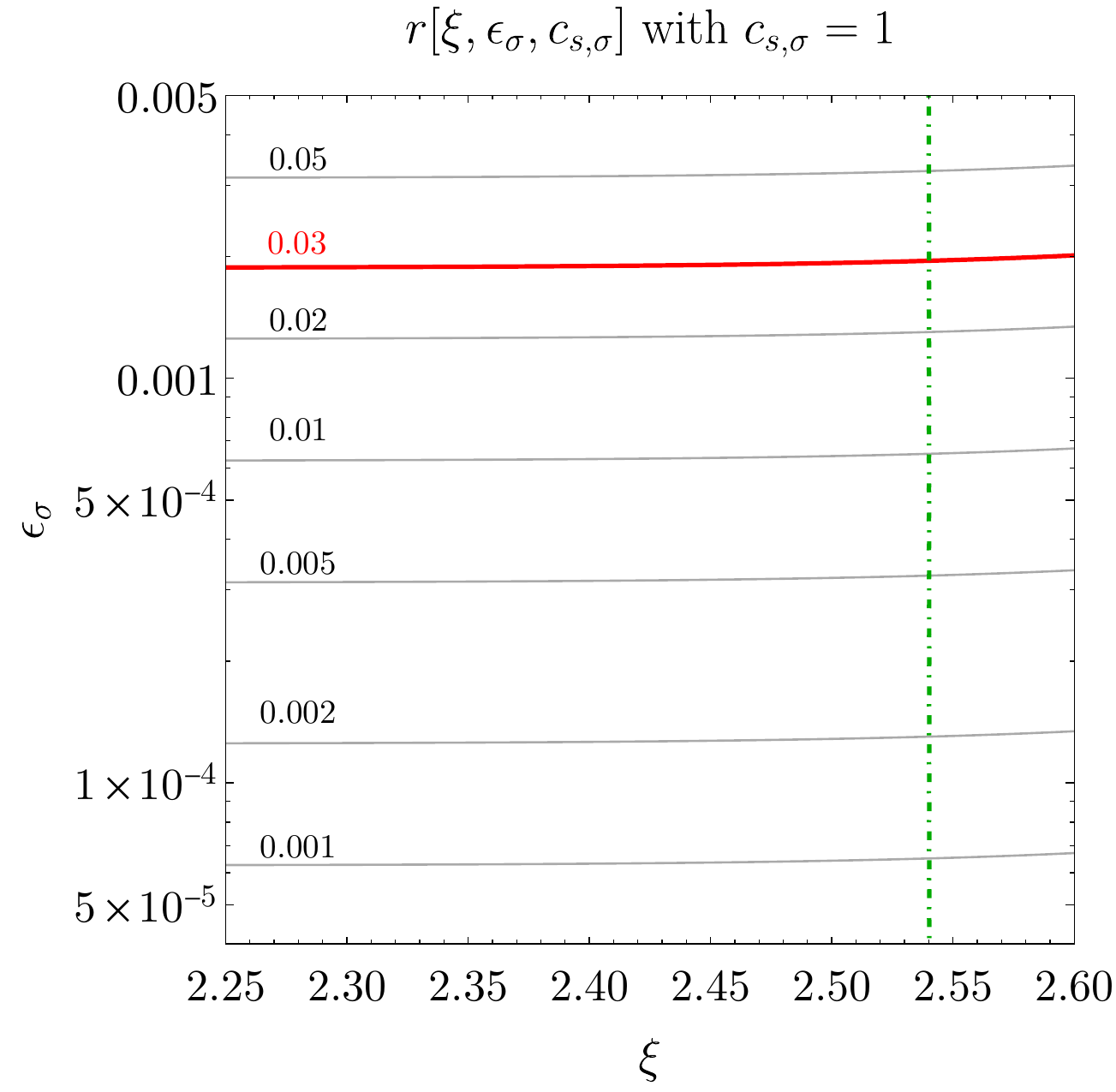}
\includegraphics[width=0.35\textwidth,angle=0]{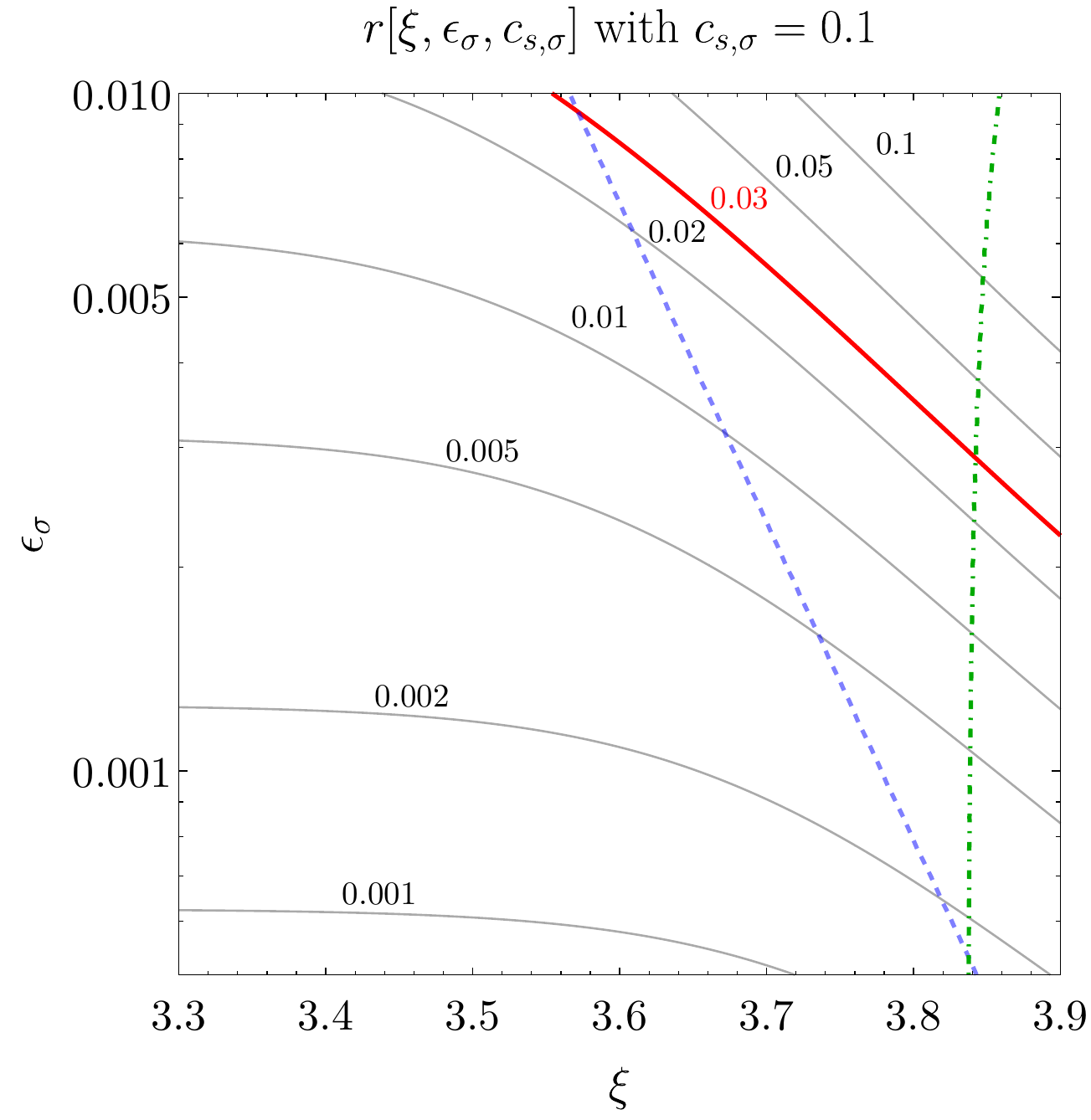}
\includegraphics[width=0.37\textwidth,angle=0]{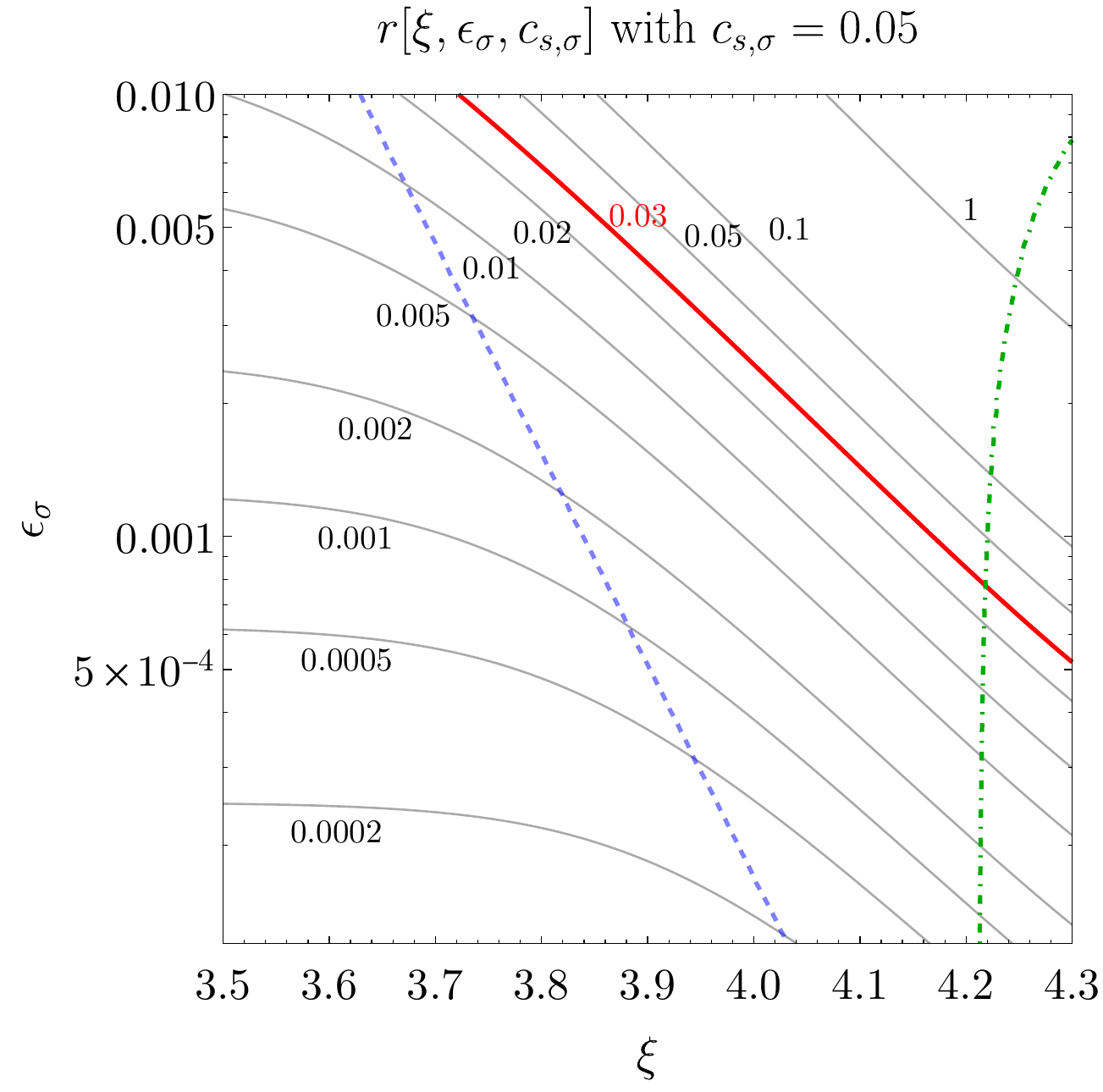}
}
\caption{
Total tensor-to-scalar-ratio $r$ as a function of $\xi$ and $\epsilon_\sigma$. The three panels correspond to three different values of the sound speed, specified on the top of the panel. The value indicated next to each line indicates the value of $r$ along that line. Values of $\xi$ on the right of the vertical green dot-dashed line lead to a too large non-Gaussianity. The area above the red line $r = 0.03$ leads to a too large tensor-to-scalar-ratio~\cite{Tristram:2021tvh,Galloni:2022mok}.
Left panel: in all the area shown the vacuum GW signal is greater than the sourced one. Central panel: in the area to the left (resp. right) of the blue dashed line the vacuum (resp. sourced) GW signal is dominant. Right panel: the area where the sourced GW signal is dominant and where the limits on non-Gaussianity and $r$ are respected, becomes larger. 
}
\label{fig:rtot}
\end{figure}

In Fig.~\ref{fig:rtot} we plot this ratio in the $\xi-\epsilon_{\sigma}$ plane. The figure consists of three panels, characterized by a progressively smaller value of $c_{s,\sigma}$. As a reference, the left panel shows the standard $c_{s,\sigma} = 1$ case. We note that in this panel the lines of equal $r$ slightly bend upwards as $\xi$ increases. This means that, for any fixed $\epsilon_\sigma$, the gauge field amplification results in a decrease of the tensor-to-scalar ratio. Moreover, in all the area shown in this panel the sourced GW power spectrum is much smaller than the vacuum one. The source GW signal is dominant only for larger values of $\xi$ than those shown here, which are ruled out by the limits on scalar non-Gaussianity. As already remarked in Ref.~\cite{Barnaby:2011vw}, the limits on scalar non-Gaussianity imply that the sourced GW anyways cannot be observed. 

The central panel is instead characterized by $c_{s,\sigma}  = 0.1$. In this panel the lines of equal $r$ bend downward. In this regime, where $\xi$ can be large enough while satisfying $P_{\zeta} \simeq {\cal P}_{\zeta}^{(0)}$, the total tensor-to-scalar ratio can be approximated as
\begin{equation}
r \simeq r_{\rm vac} + 5.4 \times 10^{-4}\epsilon_{\sigma}^2c_{s,\sigma}^2 P_{\zeta}\frac{e^{4\pi\xi}}{\xi^6}. 
\end{equation}
This means that, for any fixed $\epsilon_\sigma$, the gauge field amplification results in an increase of the tensor-to-scalar ratio. 
Another signature of this is the fact that there is now a region of parameters, situated between the (blue) dashed line and (green) vertical dot-dashed line, which is compatible with the non-Gaussianity limit and for which the sourced GW signal dominates over the vacuum one. 
In the right panel, we show the case of $c_{s,\sigma}  = 0.05$ where such observationally interesting region is broadened.

\subsection{Two-field scenario: pseudo-scalar spectator}\label{sec:two_fields}

Finally, let us consider the case in which $\sigma$, the pseudo-scalar directly coupled to the gauge field, is not the inflaton, but rather a spectator field. Instead, $\phi$ plays the role of the inflaton and supports the final curvature perturbation $\zeta$ through Eq.~\eqref{eq:phi_to_zeta}. From Eqs.~\eqref{az2} and~\eqref{az-mix}, we find, to leading order in slow roll, the equations of motion 
\begin{align}
v_\phi'' + \left( c_{s,\phi}^2 k^2 - \frac{2}{\tau^2} \right) v_\phi &= \frac{c_{s,\phi}^2-c_{s,\sigma}^2}{c_{s,\phi} c_{s,\sigma}} \, \frac{\sqrt{\epsilon_\phi \, \epsilon_\sigma}}{\tau} v_\sigma' + \frac{c_{s,\phi}^2+2 c_{s,\sigma}^2 + 3 c_{s,\phi}^2 c_{s,\sigma}^2}{c_{s,\phi} c_{s,\sigma}} \, \frac{\sqrt{\epsilon_\phi \, \epsilon_\sigma}}{\tau^2}  v_\sigma \nonumber\\
& - \frac{1}{a c_{s,\phi}}\sqrt{\frac{\epsilon_\phi}{2}} \, \frac{1}{M_p}k^4J_{1,2}, \label{eq:two_field_phi}\\ 
v_\sigma'' + \left( c_{s,\sigma}^2 k^2 - \frac{2}{\tau^2} \right) v_\sigma &\simeq 
\frac{1}{a c_{s,\sigma}}  \frac{c_{s,\sigma}^2}{\sqrt{K_{\sigma,1}}f}k^4J_3,\label{eq:two_field_sigma}
\end{align}
where the source $J_{1,2}$ is given by Eq.~\eqref{eq:def_J}, with the sound speed replaced by $c_{s,\phi}$, and where in the last expression we only keep the leading source term $J_3$ from the direct coupling.\footnote{For sufficiently small $c_{s,\sigma}$, the source $J_{1,2}$ for the $\sigma$ field yields a non-negligible contribution in sourcing $v_{\sigma}$. 
In such a case, however, we expect that the $J_{1,2}$ contribution to $v_\sigma$ provides a subleading contribution to $v_\phi$ (and hence to $\zeta$) with respect to he $J_{1,2}$ contribution that we account for in the equation for $v_\phi$. Therefore, Eqs.~\eqref{eq:two_field_phi} and~\eqref{eq:two_field_sigma} are sufficient to reliably quantify $\zeta$.}

In the present case, $J_3$ sources $v_{\sigma}$ (isocurvature component), which  is then partially converted into the curvature perturbation through the gravitational mixing between $\phi$ and $\sigma$. To estimate this production, we note that the contribution from $J_3$ to $\zeta$ is parametrically proportional to
\begin{equation}
\frac{c_{s,\phi}}{\sqrt{2\epsilon_\phi}}\, \times \frac{{\rm Max} \left[ c_{s,\phi}^2 ,\, c_{s,\sigma}^2 \right]}{c_{s,\phi}c_{s,\sigma}} \, \sqrt{\epsilon_\phi \, \epsilon_\sigma} \, \times\frac{c_{s,\sigma}}{\sqrt{K_{\sigma,1}}} \frac{1}{f}= 
{\rm Max} \left[ c_{s,\phi}^2 ,\, c_{s,\sigma}^2 \right]\frac{\xi}{M_{\rm Pl}},\label{eq:double_cs_suppress}
\end{equation}
where in the left-hand side, the first factor comes from the normalization~\eqref{eq:phi_to_zeta}, the second factor from the gravitational mixing ($v_{\sigma}\to v_\phi$), as accounted from the first two terms on the r.h.s. of Eq.~\eqref{eq:two_field_phi}, and the third factor from the direct coupling of $v_{\sigma}$ in Eq.~\eqref{eq:two_field_sigma}.
In the final step the definition of $\xi$~\eqref{xi} has been used. 
Besides being proportional to this factor, the final amplitude of $\zeta$ is also proportional to the number of $e$-foldings $\Delta N_k$ for which the spectator field is rolling since the horizon crossing of CMB modes.
During this period, the super-horizon conversion $\delta \sigma \to \zeta$ is taking place. In the standard $c_{s,\phi} = c_{s,\sigma} =1$ case, this production results in a significant amount of non-Gaussianity, unless $\Delta N_k$ is severely constrained. Specifically, one needs to assume that the spectator $\sigma$ only rolls for very few $e$-folds after the CMB modes have been produced~\cite{Ferreira:2014zia,Namba:2015gja}. In the current case, the above scaling shows that the smallness of both sound speed suppresses the sourced power spectrum $P_{\zeta}^{(1)}$ as ${\rm Max}\left[ c_{s,\phi}^4, c_{s,\sigma}^4\right]$.

To move from this qualitative discussion to a  quantitative computation of the sourced modes, let us consider the specific example of $c_{s,\phi} = c_{s,\sigma} = c_s$ with $\xi = const$. in Eqs.~\eqref{eq:two_field_phi} and~\eqref{eq:two_field_sigma}.
It is important to stress that this suppression is effective also in the case of a varying $
\xi$, and will be in act also if, for instance, the model of~\cite{Namba:2015gja} (where $
\xi$ experiences a significant variation for an order one number of e-folds) is generalized to a non-canonical kinetic term and small sound speed. In the following, for definiteness, we perform explicit computations for the simpler case of $c_{s,\phi} = c_{s,\sigma} = c_s$ with $\xi = const$, to concretely illustrate how this suppression acts, leaving to future work a detailed quantitative computation for the case of varying $\xi$.

\begin{figure}[htbp]
\begin{center}
\includegraphics[width=0.48\textwidth,angle=0]{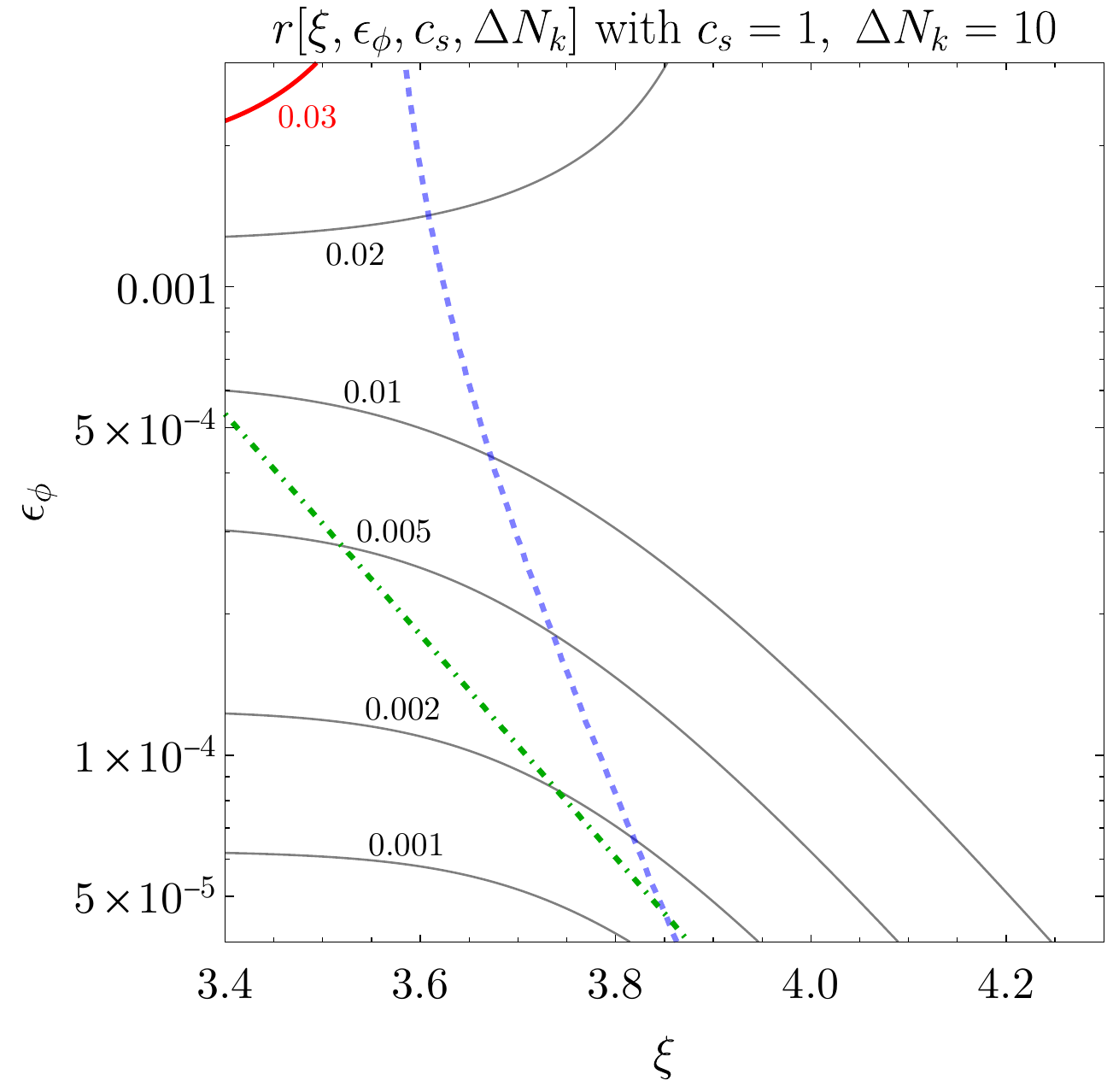}
\includegraphics[width=0.48\textwidth,angle=0]{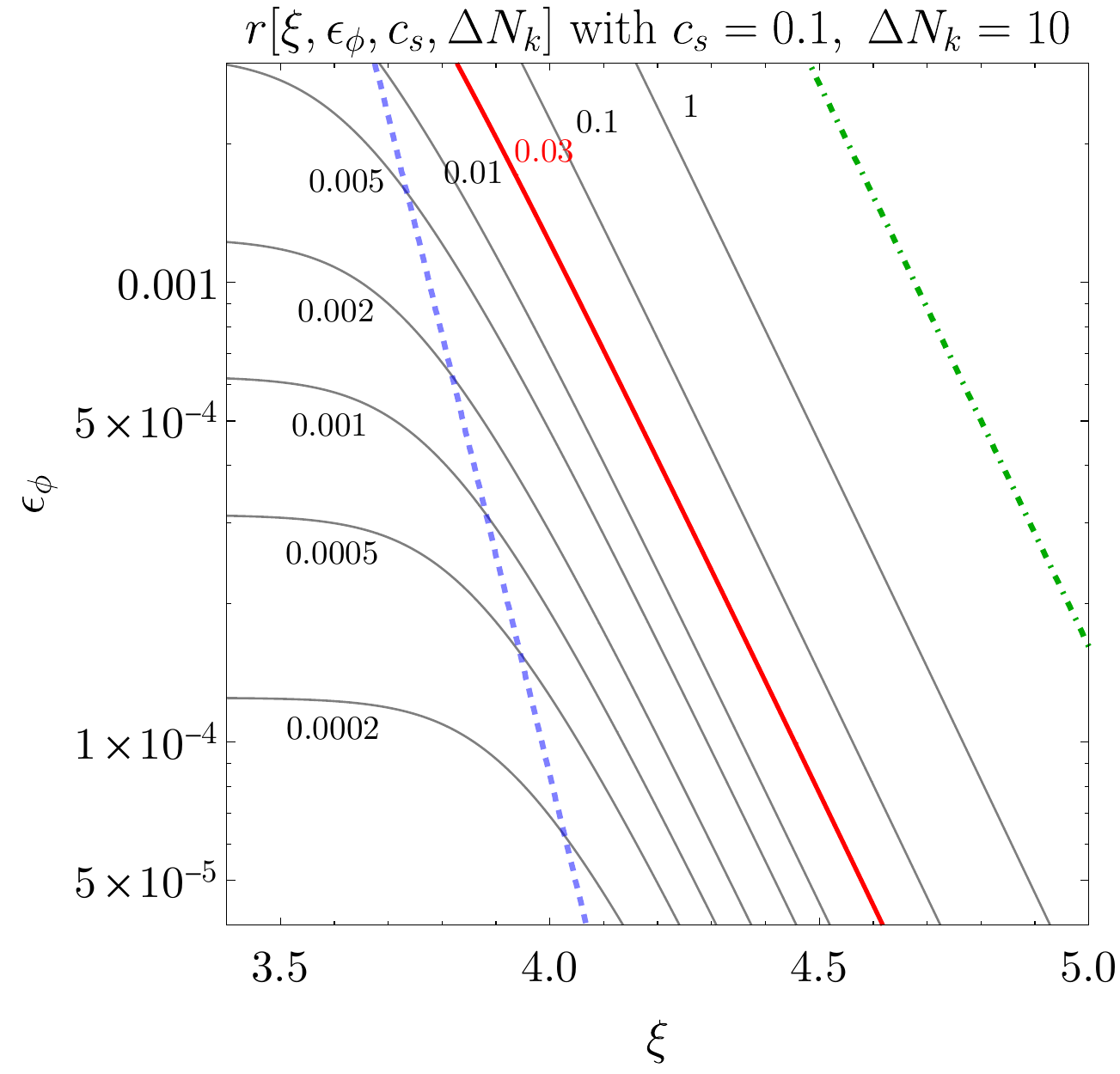}
\caption{
Analogous plot as Fig.~\ref{fig:rtot} but for the two-field scenario. The left (respectively, right) panel assumes a sound speed equal to $1$ (respectively, to $0.1$). In both panels, we assume that the spectator field rolls for $\Delta N_k  = 10$ $e$-folds after the mode at which the ratio is evaluated leaves the horizon. As in the previous figure: the red solid contour marks the current upper limit $r \simeq 0.03$; parameters above the dot-dashed green line lead to a too large scalar non-Gaussianity; the sourced tensor mode dominates over the vacuum one above the blue dashed line.
}
\label{fig:rtot_two_fields}
\end{center}
\end{figure}

In App.~\ref{app:two_field} we show that, for equal sound speeds and constant $\xi$, the equations of motion of the mode functions of the two fields can be diagonalized, and their formal solution acquires a reasonably simple form. Similarly to the single field case, we define 
\begin{equation}
\begin{aligned}
    {\cal P}_\zeta^{(1)} \left( k \right) \equiv 
    \left[\epsilon_{\phi} {\cal P}_\zeta^{(0)} \right]^2 e^{4\pi\xi} f_{2,\zeta} \left[c_s,\, \xi ,\, \Delta N_k \right],
\end{aligned}\label{eq:source_const_xi_two_fields}
\end{equation}
\begin{equation}
F\vert_{|\vec{k}_1| = |\vec{k}_2| = |\vec{k}_3| = k} \equiv \frac{\lkk \epsilon_{\phi}{\cal P}_{\zeta}^{(0)}\rkk^3}{k^6}e^{6\pi\xi}f_{3,\zeta}\left[c_{s},\, \xi ,\, \Delta N_k \right],\label{eq:f3_def_two_fields}
\end{equation}
and we evaluate the functions $f_{2,\zeta}$ and $f_{3,\zeta}$ as Eqs.~\eqref{eq:f2_full_multi} and~\eqref{eq:f3_full_multi}.
Notice that in this case, we can factor out $\epsilon_{\phi}$ as an overall factor.
Instead, $f_{2,\zeta}$ and $f_{3,\zeta}$ depend on the parameter $\Delta N_k$, which represents the number of $e$-foldings from the time the mode of interest leaves the horizon until the axion decays.
Using these equations and the relation~\eqref{eq:source_tensor} for the tensor perturbation, we plot in Fig.~\ref{fig:rtot_two_fields} the tensor-to-scalar ratio in the $\xi-\epsilon_{\phi}$ plane. We fix $\Delta N_k = 10$ and we present two different panels, for $c_s = 1$ and $c_s = 0.1$, respectively. In each panel, the contour corresponding to the current upper limit, $r \simeq 0.03$~\cite{Tristram:2021tvh,Galloni:2022mok}, is indicated with a red line. Analogously to the previous figure, we also show the 2-$\sigma$ constraint on $f_{\rm NL}^{\rm eff}$, that excludes the region above the dot-dashed green line. We also separate with a blue line the regions of parameters for which the sourced modes (above the line) or the vacuum modes (below the line) dominate the tensor signal. 

Let us start from the standard case $c_s = 1$, visualized in the left panel. As we already remarked, in this case the limits from scalar non-Gaussianity severely constrain this mechanism, and, in the allowed portion of the region shown, the tensor signal is dominated by the vacuum modes. The situation is significantly improved for $c_s = 0.1$, as shown by the right panel. In this case we find a significant portion of parameters compatible with the non-Gaussianity constraint for which the sourced tensor modes dominate over the vacuum ones. A very similar result is obtained for the case of $\Delta N_k  = 50$ (not sown in the figure), indicating that the spectator axion field is allowed to survive and roll until the end of inflation.

The marked difference between the two panels is due to the suppression of the non-Gaussian sourced scalar signal that has been discussed in Eq.~\eqref{eq:double_cs_suppress}.

\section{Discussion}\label{sec:discussion}

In this study, we revisit the phenomenology of Chern-Simons interaction between a pseudo-scalar (axion) and a U(1) gauge field during inflation, focusing on the generation of primordial perturbations.
In the standard scenario, where the axion field has a canonical kinetic term, the sourced gauge field - with an amplitude controlled by the coupling $\xi$ (proportional to the axion velocity) - sources axion perturbations $\delta \sigma$ through the inverse decay process more efficiently than it generates circularly polarized tensor perturbations. This results in a decreased total tensor-to-scalar ratio $r$. The sourced $\delta \sigma$ is highly non-Gaussian~\cite{Barnaby:2010vf}, with a shape close to the equilateral template~\cite{Barnaby:2011qe}. This results in a tight constraint on the efficiency of this mechanism~\cite{Planck:2015sxf,Planck:2015zfm,Planck:2019kim}, that prevents the observation of the sourced tensor signal for a single field scenario where the axion is identified with the inflaton~\cite{Barnaby:2010vf,Barnaby:2011qe}. This situation can be alleviated for an alternative scenario, where the axion is a spectator field and rolls only for a few $e$-folds, as the inverse decay into $\delta \sigma$ gravitationally produces curvature perturbation only during this limited period~\cite{Namba:2015gja,Ferreira:2014zia}. With a phenomenological interest, in this work we assume that the axionic inflaton (in the case of a single field), or both the axion and the inflaton (for the two-field case) have non-canonical kinetic terms, and we investigate the observational prospects for these two models, in Secs.~\ref{sec:single_field} and~\ref{sec:two_fields}, respectively.

For the axionic inflaton case, we find that
the coupling strength of Chern-Simons interaction between the canonically normalized axion degree of freedom and the vector fields attains a linear dependence on the sound speed $c_s$, see Eq.~\eqref{az-mix}. An additional $c_s$ factor arises in the relation between the curvature perturbation and the canonical axion variable, see Eq.~\eqref{eq:phi_to_zeta}. An analogous overall $c_s^2$ dependence for the amplitude of the curvature arises in the spectator scenario, as obtained in Eq.~\eqref{eq:double_cs_suppress}.
For small values of $c_s$, this leads to a strong suppression of the direct production of axion perturbation $\delta \sigma$ and the resulting curvature perturbation.

Physically, a decrease of the sound speed corresponds to an increase of the inertia of the scalar perturbation, reducing the efficiency of its production from the amplified gauge field compared to the standard case. We note that as the gauge field production and inverse decay dominantly takes place on scales beyond the sound horizon, the decrease of the sound speed affects the overall amplitude of the sourced component, but not its spectral shape (see Eq.~\eqref{Tzeta-ana} for a more detailed evaluation).

Since the coupling between the sourcing gauge field and the tensor modes do not depend on the scalar sound speed (see App.~\ref{sec:tensor} for more details), non-canonical kinetic terms leading to reduced sound speeds open up a new observational window of the model, both for the single field and the two-field scenario.
The suppression of the production of curvature perturbations allows now for i) an increase of the tensor-to-scalar ratio and ii) less stringent bounds from scalar non-Gaussianity on the amplification parameter $\xi$. As illustrated in Figs.~\ref{fig:rtot} and~\ref{fig:rtot_two_fields}, we find a region in parameter space where the sourced component dominates the tensor perturbation, with a value of $r$ detectable in the near future, while respecting the current bound on scalar non-Gaussianity. Recalling that, in the standard case, such a region does not appears for the single field scenario, and 
it is present in the spectator scenario only when the axion rolls for a very limited number of $e$-folds, allowing for non-canonical kinetic terms yields drastic change in the observational consequences of this mechanism.

We conclude our work with phenomenological implications and possible extensions of this study. In the standard single field scenario, the non-Gaussianity bound constrains the amplification parameter to $\xi \lesssim 2.5$ at the CMB scales~\cite{Barnaby:2010vf,Barnaby:2011vw,Barnaby:2011qe,Linde:2012bt,Planck:2015zfm}; however, the inflaton speed typically increases during inflation, giving rise, as we mentioned, to blue signals that can show up in smaller scales observations. See for instance Refs.~\cite{Bartolo:2016ami,LISACosmologyWorkingGroup:2024hsc} for the possible detection of this (and other) inflationary signal(s) at LISA. The spectral dependence of these signatures is very sensitive to the backreaction of the gauge fields on the axion background dynamics~\cite{Garcia-Bellido:2023ser}, on which our understanding is currently increasing thanks to improved lattice simulations~\cite{Caravano:2022epk,Figueroa:2023oxc,Caravano:2024xsb,Figueroa:2024rkr}. It will be important to investigate backreaction with non-canonical axion kinetic terms, also taking into account that in this context the amplification parameter can be greater already at CMB scales.

Another route for future development is that the underlying physics in this study - the suppression of scalar mode suppression due to the large inertia - is quite general. Therefore, once a non-canonical kinetic term is incorporated, similar conclusion would apply to other mechanisms that produce `unwanted' scalar perturbations in addition to the tensor ones. 
This may include the models discussed in the introduction (see Refs.~\cite{Watanabe:2020ctz,Dimastrogiovanni:2023oid,Murata:2024urv} for recent studies on the Chern-Simons interaction with SU(2) gauge fields~\cite{Adshead:2012kp}).

Finally, as already mentioned in the Introduction, we believe it would be extremely interesting to study whether explicit model with non-canonical axion kinetic term can be constructed from fundamental theories, in regimes which are under perturbative control. We hope to come back to this point in future work.

\section*{Acknowledgments}
The authors thank Jun'ichi Yokoyama for the fruitful discussion at the initial stage of this work, and thank Luca Martucci for helpful discussions.
J.K is supported by the JSPS Overseas Research Fellowships.
J.K and M.P acknowledge support from Istituto Nazionale di Fisica Nucleare (INFN) through the Theoretical Astroparticle Physics (TAsP) project. M.P acknowledges support from the MIUR Progetti di Ricerca di Rilevante Interesse Nazionale (PRIN) Bando 2022 - grant 20228RMX4A, funded by the European Union - Next generation EU, Mission 4, Component 1, CUP C53D23000940006.
N.B acknowledges financial support from the INFN InDark initiative and from the COSMOS network (www.cosmosnet.it) through the ASI (Italian Space Agency) Grants 2016-24-H.0, 2016-24-H.1-2018 and 2020-9-HH.0.  
N.B acknowledges support by the MUR PRIN2022 Project “BROWSEPOL: Beyond standaRd mOdel With coSmic microwavE background POLarization”-2022EJNZ53 financed by the European Union-Next Generation EU.

\appendix
\section{Computation of the sourced scalar perturbation}\label{app:scalar_detail}
Here we present the complete expressions and some details of the computation of the sourced curvature perturbations discussed in Sec.~\ref{sec:scalar_sourced}. We emphasize that our results below account for all contributions from the gauge field, including those arising from its gravitational couplings. While these interactions are much smaller than the direct Chern-Simons coupling in the standard scenario, the suppression of the latter with $c_s$ makes the former competitive at decreased sound speed, as we show in App.~\ref{app:comparison}.

\subsection{Single field case}\label{app:scalar_single}
We find the concise expression of the sourced curvature perturbation as
\begin{equation}
\begin{aligned}
& \!\!\!\!\!\!\!\! 
\hat{\zeta}^{(1)} \left( 0^- ,\, {\bm k} \right) =  
{\cal P}_\zeta^{(0)} \times 3 \sqrt{2} \pi^{7/2} \xi e^{2\pi\xi}
\Bigg[\int \frac{d^3 {\tilde p}}{\left( 2 \pi \right)^{3/2}}  \epsilon_i^{(+)} \left( {\bm p} \right)
\epsilon_i^{(+)} \left( {\hat k} - \tilde{\bm p} \right) \tilde{p}^{1/4}\left\vert \hat{k} - \tilde{{\bm p}} \right\vert^{1/4}
\\
&\quad\quad {\cal I}_\zeta \left[\epsilon_{\sigma},\, c_{s,\sigma} ,\, \xi ,\, \sqrt{{\tilde p}} , \sqrt{\left\vert \hat{k} - \tilde{{\bm p}} \right\vert}\right] \left[ {\hat a}_+ \left( {\bm p} \right) + {\hat a}_+^\dagger \left( - {\bm p} \right) \right]
\left[ {\hat a}_+ \left( {\bm k} - {\bm p} \right) + {\hat a}_+^\dagger \left( - {\bm k} + {\bm p} \right) \right] \Bigg],
\end{aligned}\label{eq:sourced_pert}
\end{equation}
which, after some algebra, results in the expression of sourced power spectrum:
\begin{equation}
\begin{aligned}
    {\cal P}_\zeta^{(1)} \left( k \right)  &= \left[ {\cal P}_\zeta^{(0)} \right]^2 \frac{9 \pi^3 \xi^2e^{4\pi\xi}}{16} \int_1^\infty dx \int_0^1 dy \frac{\left( x^2-1\right)^2}{\sqrt{x^2 - y^2}}  
{\cal I}_\zeta^2
\left[\epsilon_\sigma,\, c_{s,\sigma} ,\, \xi ,\, \sqrt{\frac{x+y}{2}}, \, \sqrt{\frac{x-y}{2}}\right]\\
&\equiv \left[ {\cal P}_\zeta^{(0)} \right]^2 e^{4\pi\xi} f_{2,\zeta} \left[\epsilon_\sigma,\,  c_{s,\sigma},\, \xi\right].
\end{aligned}\label{eq:source_const_xi_integ}
\end{equation}
Here we define the function
\begin{align}
&
{\cal I}_\zeta \left[\epsilon_{\sigma},\, c_{s,\sigma}, \, \xi ,\, \sqrt{{\tilde p}} , \sqrt{{\tilde q}}\right] \equiv 
c_{s,\sigma}^3
\left[ {\tilde p}^{1/2} + {\tilde q}^{1/2} \right]  {\cal T}_\zeta^{(E\cdot B)} \left[\xi,\, \sqrt{{\tilde p}} + \sqrt{{\tilde q}} \right] \nonumber\\
& \quad\quad + 
\epsilon_{\sigma}c_{s,\sigma}\left[ c_{s,\sigma}^2 - \left({\tilde p} - {\tilde q} \right)^2 \right] 
\left\{ 
{\cal T}_\zeta^{(E^2)} \left[\xi,\, \sqrt{{\tilde p}} + \sqrt{{\tilde q}} \right] + {\tilde p}^{1/2} {\tilde q}^{1/2}
\, {\cal T}_\zeta^{(B^2)}  \left[\xi,\, \sqrt{{\tilde p}} + \sqrt{{\tilde q}} \right] \right\} \;,\label{eq:Ifunc_single}
\end{align}
and the time integrals in terms of $x \equiv -k\tau$ as
\begin{equation}
\begin{aligned}
\!\!\!\!\!\!\!\! \!\!\!\!\!\!\!\! 
{\cal T}_\zeta^{(E^2)} \left[\xi ,\, Q \right] &\equiv  \frac{1}{3\pi^{3/2}\xi^{1/2}}
\int^{\infty}_0 dx
    \frac{-c_s x \, \cos \left( c_s x \right) + \sin \left( c_s x \right) }{c_s^3} x^{-1/2}
\exp \left[ - 2\sqrt{2\xi x} Q \right] \\
& \simeq \frac{1}{3\pi^{3/2}\xi^{1/2}}
\int^{\infty}_0 dx
    \frac{x^{5/2}}{3}
\exp \left[ - 2\sqrt{2\xi x} Q \right] =
\frac{5}{32 \sqrt{2} \pi^{3/2} \xi^4 Q^7} \;, \\
\!\!\!\!\!\!\!\! \!\!\!\!\!\!\!\! 
{\cal T}_\zeta^{(B^2)} \left[\xi ,\, Q \right] &\equiv \frac{1}{6\pi^{3/2}\xi^{3/2}}
\int^{\infty}_0 dx
 \frac{-c_s x \, \cos \left( c_s x \right) + \sin \left( c_s x \right) }{c_s^3}x^{1/2}
 \exp \left[ - 2\sqrt{2\xi x} Q \right]\\
 & \simeq \frac{1}{6\pi^{3/2}\xi^{3/2}}
\int^{\infty}_0 dx
    \frac{x^{7/2}}{3}
\exp \left[ - 2\sqrt{2\xi x} Q \right] =
\frac{35}{64 \sqrt{2} \pi^{3/2} \xi^6 Q^9} \;, \\
\!\!\!\!\!\!\!\! \!\!\!\!\!\!\!\! 
{\cal T}_\zeta^{(E\cdot B)} \left[\xi,\, Q \right] &\equiv \frac{\sqrt{2}}{3\pi^{3/2}}\int^{\infty}_0 dx
    \frac{-c_s x \, \cos \left( c_s x \right) + \sin \left( c_s x \right) }{c_s^3}
\exp \left[ - 2\sqrt{2\xi x} Q \right]\\
&\simeq \frac{\sqrt{2}}{3\pi^{3/2}}
\int^{\infty}_0 dx
    \frac{x^3}{3}
\exp \left[ - 2\sqrt{2\xi x} Q \right] = \frac{35}{64 \sqrt{2} \pi^{3/2} \xi^4 Q^8} \;,  \\
\end{aligned}
\label{Tzeta-ana}
\end{equation}
where the approximations are obtained by expanding the non exponential part in the $x \ll 1$ limit, which is appropriate for $\xi \gg1$. We note that, in this limit, the 
time integrals are independent of the sound speed, which implies that the sound speed dependence of ${\cal P}_{\zeta}^{(1)}$ arises solely from the prefactors in Eq.~\eqref{eq:Ifunc_single}. 
Consequently, when $\epsilon_{\sigma}$ is sufficiently small and the $E\cdot B$ contribution dominates, $c_{s,\sigma}$ suppresses the overall amplitude by $c_{s,\sigma}^3$ without altering the spectral shape from the standard scenario.

By performing the integral~\eqref{eq:source_const_xi_integ} for several values of parameters, specifically $\xi = 3,4,...,7$, $c_{s,\sigma} = 0.01, 0.1, 1$ and $\epsilon_{\sigma} = 0.01,0.1,1$, we find the fitting formula
\begin{equation}
\begin{aligned}
    f_{2,\zeta} \left[\epsilon_\sigma,\,  c_{s,\sigma},\, \xi\right] &\simeq 
\frac{7.47 \times 10^{-5}}{\xi^6} c_{s,\sigma}^6 
- \frac{1.92 \times 10^{-5}}{\xi^6} c_{s,\sigma}^4 \epsilon_{\sigma} 
+ \frac{4.27 \times 10^{-5}}{\xi^6} c_{s,\sigma}^6 \epsilon_{\sigma} \\
&+ \frac{1.89 \times 10^{-6}}{\xi^6} c_{s,\sigma}^2 \epsilon_{\sigma}^2 
- \frac{5.49 \times 10^{-6}}{\xi^6} c_{s,\sigma}^4 \epsilon_{\sigma}^2 
+ \frac{6.10 \times 10^{-6}}{\xi^6} c_{s,\sigma}^6 \epsilon_{\sigma}^2 \\
&+ \frac{3.49 \times 10^{-5}}{\xi^8} c_{s,\sigma}^6 \epsilon_{\sigma} 
- \frac{1.50 \times 10^{-5}}{\xi^8} c_{s,\sigma}^4 \epsilon_{\sigma} 
+ \frac{2.88 \times 10^{-6}}{\xi^8} c_{s,\sigma}^2 \epsilon_{\sigma}^2 \\
&- \frac{8.55 \times 10^{-6}}{\xi^8} c_{s,\sigma}^4 \epsilon_{\sigma}^2 
+ \frac{9.96 \times 10^{-6}}{\xi^8} c_{s,\sigma}^6 \epsilon_{\sigma}^2 
+ \frac{4.09 \times 10^{-6}}{\xi^{10}} c_{s,\sigma}^6 \epsilon_{\sigma}^2 \\
&- \frac{3.35 \times 10^{-6}}{\xi^{10}} c_{s,\sigma}^4 \epsilon_{\sigma}^2 
+ \frac{1.11 \times 10^{-6}}{\xi^{10}} c_{s,\sigma}^2 \epsilon_{\sigma}^2 \;. 
\end{aligned}\label{eq:f2_full}
\end{equation}
Note that, to precisely estimate the coefficients of the ${\cal O}(\epsilon_{\sigma}^2)$ terms, we set relatively large values for the sample points.
In Eq.~\eqref{eq:f2_full}, the first term at right hand side originates from the direct interaction, and it agrees with Eq.~(3.19) of Ref.~\cite{Barnaby:2011vw} for $c_{s,\sigma} = 1$. The remaining terms with ${\cal O}(\epsilon_{\sigma}^2)$ are due to the gravitational coupling and the ones with ${\cal O}(\epsilon_{\sigma})$ are due to the interference. We see that these terms are slow roll suppressed, but they can be competitive with the direct term at small sound speed. 

Similarly to the power spectrum, we express the three point function as 
\begin{equation}
F\lmk\vec{k}_1,\,\vec{k}_2,\,\vec{k}_3\rmk = \frac{\lkk {\cal P}_{\zeta}^{(0)}\rkk^3}{k_1^2k_2^2k_3^2}e^{6\pi\xi}f_{3,\zeta} \;, 
\end{equation}
and we then obtain
\begin{align}
f_{3,\zeta} &\left[ \epsilon_{\phi} ,\,c_s,\,\xi ,\, x_2 ,\, x_3 \right] = 2^{9/2} \, 27 \, \pi^{21/2} \xi^3 
\int \frac{d^3 {\tilde p}}{\left( 2 \pi \right)^{9/2} x_2^2 x_3^2} {\rm Re} \left\{ {\tilde P} \left[ {\tilde {\bm p}},\, {\tilde {\bm p}} + {\hat k}_1 ,\, {\tilde {\bm p}} - x_3 {\hat k}_3 \right] \right\} \nonumber\\
&  \sqrt{{\tilde p} \left\vert {\tilde {\bm p}} + {\hat k}_1 \right\vert \left\vert {\tilde {\bm p}} - x_3 {\hat k}_3 \right\vert} \sqrt{x_2x_3}  \, 
{\cal I}_\zeta \left[ \epsilon_{\phi} ,\, c_s,\, \xi ,\, \sqrt{\tilde p}, \sqrt{\left\vert {\tilde {\bm p}} + {\hat k}_1 \right\vert} \right]  \nonumber\\
& {\cal I}_\zeta \left[\epsilon_{\phi} ,\, c_s,\, \xi ,\, \frac{\sqrt{\left\vert {\tilde {\bm p}} + {\hat k}_1 \right\vert}}{\sqrt{x_2}}+\frac{\sqrt{\left\vert {\tilde {\bm p}} - x_3 {\hat k}_3 \right\vert} }{\sqrt{x_2}}  \right] 
{\cal I}_\zeta \left[ \epsilon_{\phi} ,\, c_s,\, \xi ,\, \frac{\sqrt{\left\vert {\tilde {\bm p}} - x_3 {\hat k}_3 \right\vert}}{\sqrt{x_3}} + \frac{\sqrt{\tilde p}}{\sqrt{x_3}} \right],
\label{f3zeta-final}
\end{align}
where $x_{2,3} \equiv k_{2,3} / k_1$, and where we introduce
\begin{equation}
{\tilde P} \left[ \vec{v}_1 ,\, \vec{v}_2 ,\, \vec{v}_3 \right] \equiv \epsilon_i^{(+)*} \left( \vec{v}_1 \right) \epsilon_i^{(+)} \left( \vec{v}_2 \right) \epsilon_j^{(+)*} \left( \vec{v}_2 \right) \epsilon_j^{(+)} \left( \vec{v}_3 \right) \epsilon_k^{(+)*} \left( \vec{v}_3 \right) \epsilon_k^{(+)} \left( \vec{v}_1 \right).
\end{equation}

For the exact equilateral configuration $x_2 = x_3 = 1$, we find that $f_{3,\zeta}$ is well fitted by
\begin{equation}
\begin{aligned}
f_{3,\zeta}\left[\epsilon_\sigma,\,  c_{s,\sigma},\, \xi\right] &\simeq
\frac{2.46 \times 10^{-5}}{\xi^9} c_{s,\sigma}^9 
- \frac{8.19 \times 10^{-6}}{\xi^9} c_{s,\sigma}^7 \epsilon_{\sigma} 
+ \frac{2.10 \times 10^{-5}}{\xi^9} c_{s,\sigma}^9 \epsilon_{\sigma} \\
& + \frac{7.39 \times 10^{-7}}{\xi^9} c_{s,\sigma}^5 \epsilon_{\sigma}^2 
+ \frac{1.42 \times 10^{-6}}{\xi^9} c_{s,\sigma}^7 \epsilon_{\sigma}^2 
+ \frac{1.43 \times 10^{-6}}{\xi^9} c_{s,\sigma}^9 \epsilon_{\sigma}^2 \\
& - \frac{4.78 \times 10^{-9}}{\xi^9} c_{s,\sigma}^3 \epsilon_{\sigma}^3 
+ \frac{2.28 \times 10^{-7}}{\xi^9} c_{s,\sigma}^5 \epsilon_{\sigma}^3 
- \frac{6.87 \times 10^{-7}}{\xi^9} c_{s,\sigma}^7 \epsilon_{\sigma}^3 \\
& - \frac{6.96 \times 10^{-7}}{\xi^9} c_{s,\sigma}^9 \epsilon_{\sigma}^3 
+ \frac{1.12 \times 10^{-6}}{\xi^{11}} c_{s,\sigma}^5 \epsilon_{\sigma}^2 
+ \frac{1.27 \times 10^{-6}}{\xi^{11}} c_{s,\sigma}^7 \epsilon_{\sigma}^2 \\
& + \frac{1.27 \times 10^{-6}}{\xi^{11}} c_{s,\sigma}^9 \epsilon_{\sigma}^2 
- \frac{1.19 \times 10^{-8}}{\xi^{11}} c_{s,\sigma}^3 \epsilon_{\sigma}^3 
+ \frac{5.25 \times 10^{-7}}{\xi^{11}} c_{s,\sigma}^5 \epsilon_{\sigma}^3 \\
& - \frac{8.73 \times 10^{-8}}{\xi^{11}} c_{s,\sigma}^7 \epsilon_{\sigma}^3 
- \frac{9.34 \times 10^{-8}}{\xi^{11}} c_{s,\sigma}^9 \epsilon_{\sigma}^3 
- \frac{6.30 \times 10^{-6}}{\xi^{11}} c_{s,\sigma}^7 \epsilon_{\sigma} \\
& + \frac{1.73 \times 10^{-5}}{\xi^{11}} c_{s,\sigma}^9 \epsilon_{\sigma} 
+ \frac{4.24 \times 10^{-7}}{\xi^{13}} c_{s,\sigma}^5 \epsilon_{\sigma}^2 
+ \frac{5.66 \times 10^{-7}}{\xi^{13}} c_{s,\sigma}^7 \epsilon_{\sigma}^2 \\
& + \frac{5.68 \times 10^{-7}}{\xi^{13}} c_{s,\sigma}^9 \epsilon_{\sigma}^2 
- \frac{1.01 \times 10^{-8}}{\xi^{13}} c_{s,\sigma}^3 \epsilon_{\sigma}^3 
+ \frac{4.38 \times 10^{-7}}{\xi^{13}} c_{s,\sigma}^5 \epsilon_{\sigma}^3 \\
& - \frac{6.85 \times 10^{-8}}{\xi^{13}} c_{s,\sigma}^7 \epsilon_{\sigma}^3 
- \frac{7.36 \times 10^{-8}}{\xi^{13}} c_{s,\sigma}^9 \epsilon_{\sigma}^3 
- \frac{1.84 \times 10^{-9}}{\xi^{15}} c_{s,\sigma}^3 \epsilon_{\sigma}^3 \\
& + \frac{9.92 \times 10^{-9}}{\xi^{15}} c_{s,\sigma}^5 \epsilon_{\sigma}^3 
+ \frac{3.18 \times 10^{-8}}{\xi^{15}} c_{s,\sigma}^7 \epsilon_{\sigma}^3 
+ \frac{3.21 \times 10^{-8}}{\xi^{15}} c_{s,\sigma}^9 \epsilon_{\sigma}^3,
\end{aligned}\label{eq:f3_full}
\end{equation}
where we assumed the same parameter range as the two-point function.

\subsection{Two fields with the equal sound speed}\label{app:two_field}
Let us consider the two-field case with equal sound speeds, as this greatly simplifies their gravitational mixing. The equations of motion for the canonical variables are
\begin{align}
\left( \partial_{\tau}^2 +  c_s^2 k^2 - \frac{M_{IJ}}{\tau^2} \right) v_I  = S_I,
\end{align}
where the source terms are given by 
\begin{align}
    S_{\phi} \equiv - \frac{1}{a c_s}\sqrt{\frac{\epsilon_\sigma}{2}} \, \frac{1}{M_p} 
k^4 J_{1,2}
,\quad\quad
S_{\sigma} \equiv \frac{1}{a c_s}  \frac{c_s^2}{\sqrt{K_{\sigma,1}}f}k^4J_3,
\end{align}
with $J_{1,2,3}$ introduced in Eq.~(\ref{eq:def_J}) of the main text.

To leading order in slow roll, the mass matrix can be written as $M_{ij} \equiv -\tau^2\tilde{M}_{ij}$, with 
\begin{equation}
\begin{aligned}
    M_{IJ} &= 
    \begin{pmatrix}
        2 + \frac{3}{2}(5 + c_s^2)\, \epsilon_{\phi} + \frac{3}{2}(3 - c_s^2)\, \epsilon_{\sigma} - \frac{3}{2}(1 + c_s^2)\, \eta_{\phi} 
        & \quad 3(1 + c_s^2)\, \sqrt{\epsilon_{\phi}\epsilon_{\sigma}} \\[8pt]
        3(1 + c_s^2)\, \sqrt{\epsilon_{\phi}\epsilon_{\sigma}} 
        & \quad 2 + \frac{3}{2}(5 + c_s^2)\, \epsilon_{\sigma} + \frac{3}{2}(3 - c_s^2)\, \epsilon_{\phi} - \frac{3}{2}(1 + c_s^2)\, \eta_{\sigma}
    \end{pmatrix},
\end{aligned}
\end{equation}
where we recover the mixing term in Eq.~\eqref{eq:two_field_sigma}. We denote the rotation matrix for the diagonalization of this matrix, and the eigenvalues, as
\begin{equation}
M = C^T \Lambda C, \quad
C = \begin{pmatrix}
\cos \theta & \sin \theta \\
-\sin \theta & \cos \theta
\end{pmatrix}, \quad
\Lambda = \begin{pmatrix}
\lambda_\phi & 0 \\
0 & \lambda_\sigma
\end{pmatrix}.
\end{equation}
The diagonalization results in the two decoupled equations for the eigenmodes $\psi_i \equiv C_{ij}v_J$ as
\begin{align}
\left( \partial_{\tau}^2 +  c_s^2 k^2 - \frac{\lambda_I}{\tau^2} \right) \psi_I  = C_{IJ}S_J.
\end{align}
The formal solutions is obtained as
\begin{equation}
\begin{aligned}
    v_{\phi}\left( \tau ,\, \vec{k} \right)  = C^{-1}_{\phi J}\psi_J
    &= \int_{-\infty}^{\tau} d\tau^{\prime} 
    \left\{
    \lmk
    \cos^2\theta 
    \tilde{G}_{c_sk}^{\lambda_{\phi}}(\tau,\tau^{\prime})
    + \sin^2\theta \tilde{G}_{c_sk}^{\lambda_{\sigma}}(\tau,\tau^{\prime})
    \rmk S_{\phi}(\tau^{\prime}) \right.\\
    & \quad\quad\quad\quad 
    +\left. 
    \sin\theta\cos\theta
    \lmk  
    \tilde{G}_{c_sk}^{\lambda_{\phi}}(\tau,\tau^{\prime}) - \tilde{G}_{c_sk}^{\lambda_{\sigma}}(\tau,\tau^{\prime})
    \rmk 
    S_{\sigma}(\tau^{\prime})
    \right\},
\end{aligned}
\end{equation}
where $ \tilde{G}_{c_sk}^{\lambda_{\phi}}$ and $ \tilde{G}_{c_sk}^{\lambda_{\sigma}}$ are the Green functions of the diagonalized equations. By expanding this solution in slow roll, we find the leading contributions as 
\begin{equation}
\begin{aligned}
    v_{\phi}\left( \tau ,\, \vec{k} \right)  &\simeq \int_{-\infty}^{\tau} d\tau^{\prime} 
    \tilde{G}_{c_sk}(\tau,\tau^{\prime})
    \left\{
    S_{\phi}(\tau^{\prime}) +
    \frac{\delta \lambda_{\phi} - \delta \lambda_{\sigma}}{3}\sin\theta\cos\theta\log\lmk\frac{\tau^{\prime}}{\tau}\rmk
    S_{\sigma}(\tau^{\prime})
    \right\}\\
    &= \int_{-\infty}^{\tau} d\tau^{\prime} 
    \tilde{G}_{c_sk}(\tau,\tau^{\prime})
    \left\{
    S_{\phi}(\tau^{\prime}) +
    (1 + c_s^2)\sqrt{\epsilon_{\phi}\epsilon_{\sigma}}\log\lmk\frac{\tau^{\prime}}{\tau}\rmk
    S_{\sigma}(\tau^{\prime})
    \right\},
\end{aligned}
\end{equation}
In the super-horizon limit $|k\tau| \ll 1$, the sourced curvature perturbation in terms of the rescaled momentum can be expressed as
\begin{equation}
\begin{aligned}
    \hat{\zeta}^{(1)}\left(x \ll -1,\, \vec{k} \right)  \simeq
\frac{H^2}{M_p^2} \, \int^{\infty}_{x} d x' \;
\frac{x^{\prime 3}}{3}
\left[ \frac{1}{2} J_{1,2} - 
\log\lmk\frac{x^{\prime}}{x}\rmk\xi_\sigma \, c_s^2(1 + c_s^2) J_3 \right]. 
\label{eq:zeta_two_fields}
\end{aligned}
\end{equation}
Notice that while the contribution from $S_{\phi}(\tau)$ converge in the limit of $\tau \to 0$, that from $S_{\sigma}(\tau)$ continues to grow, which is interpreted as the super-horizon sourcing of the curvature mode from the isocurvature one~\cite{Namba:2015gja}. We assume that, $\Delta N_k$ $e$-foldings after the mode $k$ crossed the horizon, the axion reaches the minimum of its potential, and its energy is quickly redshifted, thus posing an end to the sourcing. Then, the contribution from $S_{\sigma}(\tau)$ is $\propto \Delta N_k$, and this term dominates over the $J_{1,2}$ contribution in the standard $c_s = 1$ case. 

As can be seen from the similarity between Eqs.~\eqref{eq:zeta_single_cs_1} and~\eqref{eq:zeta_two_fields}, the sourced curvature power spectrum can be expressed similarly to Eq.~\eqref{eq:source_const_xi} with a minimal modification:
\begin{equation}
\begin{aligned}
    {\cal P}_\zeta^{(1)} \left( k \right)  &= \left[ \epsilon_\phi \, {\cal P}_\zeta^{(0)} \right]^2 \frac{9 \pi^3 \xi^2e^{4\pi\xi}}{16} \int_1^\infty dx \int_0^1 dy \frac{\left( x^2-1\right)^2}{\sqrt{x^2 - y^2}}  
{\cal I}_\zeta^2 \left[c_s ,\, \xi ,\, \sqrt{\frac{x+y}{2}}, \, \sqrt{\frac{x-y}{2}} ;\, \Delta N_k \right]\\
&\equiv \left[ \epsilon_\phi \, {\cal P}_\zeta^{(0)} \right]^2 e^{4\pi\xi} f_{2,\zeta} \left[ c_s,\, \xi ,\, \Delta N_k \right] \;,
\end{aligned}\label{eq:source_const_xi_2}
\end{equation}
where we defined the function
\begin{align}
&
{\cal I}_\zeta \left[ c_s, \, \xi ,\, \sqrt{{\tilde p}} , \sqrt{{\tilde q}} ;\, \Delta N_k \right] \equiv 
c_s^3(1 + c_s^2)
\left[ {\tilde p}^{1/2} + {\tilde q}^{1/2} \right]  {\cal T}_\zeta^{(E\cdot B)} \left[\xi,\, \sqrt{{\tilde p}} + \sqrt{{\tilde q}} ;\, \Delta N_k \right] \nonumber\\
& \quad\quad + c_s\left[ c_s^2 - \left( {\tilde p} - {\tilde q} \right)^2 \right] 
\left\{ 
{\cal T}_\zeta^{(E^2)} \left[\xi,\, \sqrt{{\tilde p}} + \sqrt{{\tilde q}} ;\, \Delta N_k \right] + {\tilde p}^{1/2} {\tilde q}^{1/2}
\, {\cal T}_\zeta^{(B^2)}  \left[\xi,\, \sqrt{{\tilde p}} + \sqrt{{\tilde q}} ;\, \Delta N_k \right] \right\} \;, \label{eq:Ifunc_for_two_diag}
\end{align}
and the time integrals
\begin{align}
\!\!\!\!\!\!\!\! \!\!\!\!\!\!\!\! 
{\cal T}_\zeta^{(E\cdot B)} \left[\xi,\, Q ;\, \Delta N_k \right] &\equiv \frac{\sqrt{2}}{3\pi^{3/2}}\int^{\infty}_{e^{-\Delta N_k}} dx^{\prime}
    \lmk \log x^{\prime} + \Delta N_k \rmk \frac{x^{\prime 3}}{3} 
\exp \left[ - 2\sqrt{2\xi x^{\prime}} Q \right],\nonumber\\
\!\!\!\!\!\!\!\! \!\!\!\!\!\!\!\! 
{\cal T}_\zeta^{(E^2)} \left[\xi ,\, Q ;\, \Delta N_k\right] &\equiv  \frac{1}{3\pi^{3/2}\xi^{1/2}}
\int^{\infty}_{e^{-\Delta N_k}} dx^{\prime}
    \frac{x^{\prime 5/2}}{3} 
\exp \left[ - 2\sqrt{2\xi x^{\prime}} Q \right], \nonumber\\
\!\!\!\!\!\!\!\! \!\!\!\!\!\!\!\! 
{\cal T}_\zeta^{(B^2)} \left[\xi ,\, Q ;\, \Delta N_k\right] &\equiv \frac{1}{6\pi^{3/2}\xi^{3/2}}
\int^{\infty}_{e^{-\Delta N_k}} dx^{\prime}
 \frac{x^{\prime 7/2}}{3} 
 \exp \left[ - 2\sqrt{2\xi x^{\prime}} Q \right].
\end{align}
Notice that in Eq.~\eqref{eq:source_const_xi_2} we factor out $\epsilon_{\phi}$ since the $J_3$ contribution is also proportional to the slow-roll parameter, due to the gravitational mixing. In the same spirit, we express the 3-point function as
\begin{equation}
F\lmk\vec{k}_1,\,\vec{k}_2,\,\vec{k}_3\rmk =
\frac{\left[ \epsilon_\phi \, {\cal P}_\zeta^{(0)} \right]^3}{k_1^2 k_2^2 k_3^2} e^{6\pi\xi} f_{3,\zeta} \;,
\end{equation}
to obtain 
\begin{align}
f_{3,\zeta} &\left[ \,c_s,\,\xi ,\, x_2 ,\, x_3 ;\, \Delta N_k\right] = 2^{9/2} \, 27 \, \pi^{21/2} \xi^3 
\int \frac{d^3 {\tilde p}}{\left( 2 \pi \right)^{9/2} x_2^2 x_3^2} {\rm Re} \left\{ {\tilde P} \left[ {\tilde {\bm p}},\, {\tilde {\bm p}} + {\hat k}_1 ,\, {\tilde {\bm p}} - x_3 {\hat k}_3 \right] \right\} \nonumber\\
&  \sqrt{{\tilde p} \left\vert {\tilde {\bm p}} + {\hat k}_1 \right\vert \left\vert {\tilde {\bm p}} - x_3 {\hat k}_3 \right\vert} \sqrt{x_2x_3}  \, 
{\cal I}_\zeta \left[ c_s,\, \xi ,\, \sqrt{\tilde p}, \sqrt{\left\vert {\tilde {\bm p}} + {\hat k}_1 \right\vert} ;\, \Delta N_k \right]  \nonumber\\
& {\cal I}_\zeta \left[c_s,\, \xi ,\, \frac{\sqrt{\left\vert {\tilde {\bm p}} + {\hat k}_1 \right\vert}}{\sqrt{x_2}}+\frac{\sqrt{\left\vert {\tilde {\bm p}} - x_3 {\hat k}_3 \right\vert} }{\sqrt{x_2}} ;\, \Delta N_k  \right] 
{\cal I}_\zeta \left[c_s,\, \xi ,\, \frac{\sqrt{\left\vert {\tilde {\bm p}} - x_3 {\hat k}_3 \right\vert}}{\sqrt{x_3}} + \frac{\sqrt{\tilde p}}{\sqrt{x_3}};\, \Delta N_k \right].
\label{f3zeta-final_two_fields}
\end{align}
For large $\xi$ and $\Delta N_k \gtrsim {\cal O}(1)$, the time integrals are well approximated by
\begin{equation}
\!\!\!\!\!\!\!\! \!\!\!\!\!\!\!\! 
{\cal T}_\zeta^{(E^2)} \simeq \frac{5}{32 \sqrt{2} \pi^{3/2} \xi^4 Q^7} \;\;,\;\; 
{\cal T}_\zeta^{(B^2)} \simeq \frac{35}{64 \sqrt{2} \pi^{3/2} \xi^6 Q^9} \;\;,\;\; 
{\cal T}_\zeta^{(E\cdot B)} \simeq  \frac{35\Delta N_k}{64 \sqrt{2} \pi^{3/2} \xi^4 Q^8} \;. \\
\end{equation}
By comparing with Eq.~\eqref{Tzeta-ana}, $\Delta N_k$ dependence appears only in the $E\cdot B$ contribution as a proportional factor.
In addition, the $E\cdot B$ contribution in Eqs.~\eqref{eq:Ifunc_single} and~\eqref{eq:Ifunc_for_two_diag} differs by a factor $(1 + c_s^2)$.
Therefore, by multiplying $\Delta N_k(1+c_s^2)$ the same number of times as the terms originating from $E\cdot B$ and setting $\epsilon_{\sigma} = 1$ in Eqs.~\eqref{eq:f2_full} and~\eqref{eq:f3_full}, we can obtain the fitting functions in the present case. In the same range of parameters as before, they are given by
\begin{equation}
\begin{aligned}
    f_{2,\zeta}
    &\simeq 
    \frac{7.47 \times 10^{-5}}{\xi^6} c_s^6(1+c_s^2)^2\Delta N_k^2 
- \frac{1.92 \times 10^{-5}}{\xi^6} c_s^4 (1+c_s^2)\Delta N_k
+ \frac{4.27 \times 10^{-5}}{\xi^6} c_s^6 (1+c_s^2)\Delta N_k \\
&+ \frac{1.89 \times 10^{-6}}{\xi^6} c_s^2 
- \frac{5.49 \times 10^{-6}}{\xi^6} c_s^4 
+ \frac{6.10 \times 10^{-6}}{\xi^6} c_s^6 \\
&+ \frac{3.49 \times 10^{-5}}{\xi^8} c_s^6 (1+c_s^2)\Delta N_k
- \frac{1.50 \times 10^{-5}}{\xi^8} c_s^4 (1+c_s^2)\Delta N_k
+ \frac{2.88 \times 10^{-6}}{\xi^8} c_s^2  \\
&- \frac{8.55 \times 10^{-6}}{\xi^8} c_s^4 
+ \frac{9.96 \times 10^{-6}}{\xi^8} c_s^6 
+ \frac{4.09 \times 10^{-6}}{\xi^{10}} c_s^6 \\
&- \frac{3.35 \times 10^{-6}}{\xi^{10}} c_s^4 
+ \frac{1.11 \times 10^{-6}}{\xi^{10}} c_s^2 ,
\end{aligned}\label{eq:f2_full_multi}
\end{equation}
and by
\begin{equation}
\begin{aligned}
f_{3,\zeta} 
&\simeq 
\frac{2.46 \times 10^{-5}}{\xi^9} c_s^9
(1+c_s^2)^3\Delta N_k^3
- \frac{8.19 \times 10^{-6}}{\xi^9} c_s^7 (1+c_s^2)^2\Delta N_k^2 
+ \frac{2.10 \times 10^{-5}}{\xi^9} c_s^9 (1+c_s^2)^2\Delta N_k^2 \\
& + \frac{7.39 \times 10^{-7}}{\xi^9} c_s^5 (1+c_s^2)\Delta N_k 
+ \frac{1.42 \times 10^{-6}}{\xi^9} c_s^7 (1+c_s^2)\Delta N_k 
+ \frac{1.43 \times 10^{-6}}{\xi^9} c_s^9 (1+c_s^2)\Delta N_k \\
& - \frac{4.78 \times 10^{-9}}{\xi^9} c_s^3 
+ \frac{2.28 \times 10^{-7}}{\xi^9} c_s^5 
- \frac{6.87 \times 10^{-7}}{\xi^9} c_s^7\\
& - \frac{6.96 \times 10^{-7}}{\xi^9} c_s^9
+ \frac{1.12 \times 10^{-6}}{\xi^{11}} c_s^5 (1+c_s^2)\Delta N_k 
+ \frac{1.27 \times 10^{-6}}{\xi^{11}} c_s^7 (1+c_s^2)\Delta N_k \\
& + \frac{1.27 \times 10^{-6}}{\xi^{11}} c_s^9 (1+c_s^2)\Delta N_k 
- \frac{1.19 \times 10^{-8}}{\xi^{11}} c_s^3
+ \frac{5.25 \times 10^{-7}}{\xi^{11}} c_s^5\\
& - \frac{8.73 \times 10^{-8}}{\xi^{11}} c_s^7
- \frac{9.34 \times 10^{-8}}{\xi^{11}} c_s^9
- \frac{6.30 \times 10^{-6}}{\xi^{11}} c_s^7 (1+c_s^2)^2 \Delta N_k^2 \\
& + \frac{1.73 \times 10^{-5}}{\xi^{11}} c_s^9 (1+c_s^2)^2 \Delta N_k^2 
+ \frac{4.24 \times 10^{-7}}{\xi^{13}} c_s^5 (1+c_s^2)\Delta N_k 
+ \frac{5.66 \times 10^{-7}}{\xi^{13}} c_s^7 (1+c_s^2)\Delta N_k \\
& + \frac{5.68 \times 10^{-7}}{\xi^{13}} c_s^9 (1+c_s^2)\Delta N_k 
- \frac{1.01 \times 10^{-8}}{\xi^{13}} c_s^3
+ \frac{4.38 \times 10^{-7}}{\xi^{13}} c_s^5 \\
& - \frac{6.85 \times 10^{-8}}{\xi^{13}} c_s^7
- \frac{7.36 \times 10^{-8}}{\xi^{13}} c_s^9
- \frac{1.84 \times 10^{-9}}{\xi^{15}} c_s^3\\
& + \frac{9.92 \times 10^{-9}}{\xi^{15}} c_s^5
+ \frac{3.18 \times 10^{-8}}{\xi^{15}} c_s^7
+ \frac{3.21 \times 10^{-8}}{\xi^{15}} c_s^9.
\end{aligned}\label{eq:f3_full_multi}
\end{equation}
Let us note that, due to the absence of relative slow-roll suppression (which was instead existing in the single field case), the contribution from $J_{1,2}$ (leading to the terms that do not depend on $\Delta N_k$) can now become even more relevant at small $c_s$ with respect to what it did in the single field case.

\section{Sourced tensor perturbation}\label{sec:tensor}

In this appendix we provide some details on the tensor sector. The tensor perturbation is introduced as 
\begin{equation}
    g^{ij} = \frac{1}{a^2} \left( \delta_{ij} - h_{ij}^{TT} \right).
\end{equation}

The gauge field kinetic term in Eq.~\eqref{eq:lag_mat} yields the interaction term
\begin{equation}
\Delta\mathcal{L}_A = - a^4 \left[ \frac{1}{2} h_{ij}^{TT} \left( \hat{E}_i \hat{E}_j + \hat{B}_i \hat{B}_j \right)
\right].
\label{AA-tensor}
\end{equation}
Including the kinetic terms and redefining $h_{ij} \equiv \frac{2}{M_p a} H_{ij}$ we have the total action for the tensor modes 
\begin{equation}
S = \int d^4 x  \left[ \frac{1}{2} H_{ij}' H_{ij}' - \frac{1}{2} H_{ij,k} H_{ij,k} + \frac{a''}{2 a} H_{ij} H_{ij} - \frac{a^3}{M_p} H_{ij} \left( \hat{E}_i \hat{E}_j + \hat{B}_i \hat{B}_j \right) \right].
\end{equation}
We decompose the canonical variable as
\begin{equation}
H_{ij} \left( \tau ,\, {\bm x} \right) = \int \frac{d^3 k}{\left( 2 \pi \right)^{3/2}} {\rm e}^{i {\bm k} \cdot {\bm x}} \sum_\lambda \Pi_{ij,\lambda}^* \left( {\hat k} \right) Q_\lambda \left( \tau ,\, {\bm k} \right) \;, 
\end{equation}
where $\Pi_{ij,\lambda}^* \left( {\hat k} \right) \equiv \epsilon_i^{(\lambda)} \left( {\hat k} \right) \epsilon_j^{(\lambda)} \left( {\hat k} \right)$. Extremizing the above action, it is immediate to see that the canonical variable obeys the equation~\cite{Namba:2015gja} 
\begin{equation}
\left( \frac{\partial^2}{\partial \tau^2} + k^2 - \frac{a''}{a} \right) Q_\lambda \left( \tau ,\, {\bm k} \right) = - \frac{a^3}{M_p} \Pi_{ij,\lambda} \left( {\hat k} \right) \int \frac{d^3 x}{\left( 2 \pi \right)^{3/2}} {\rm e}^{-i {\bm k} \cdot {\bm x}} \left( \hat{E}_i \hat{E}_j + \hat{B}_i \hat{B}_j \right) \equiv {\cal S}_\lambda \left( \tau ,\, {\bm k} \right) \;, 
\end{equation}
which can be decomposed into the homogeneous solution and the particular solution as
\begin{align}
Q_\lambda \left( \tau ,\, {\bm k} \right) &= Q_\lambda^{(0)} \left( \tau ,\, {\bm k} \right) + Q_\lambda^{(s)} \left( \tau ,\, {\bm k} \right) \;, \nonumber\\ 
Q_\lambda^{(0)} \left( \tau ,\, {\bm k} \right) &= h_\lambda \left( \tau ,\, k \right) {\hat a}_\lambda \left( {\bm k} \right) + {\rm h.c.} \;\;,\;\; h_\lambda \left( \tau ,\, k \right) = \frac{{\rm e}^{-i k \tau}}{\sqrt{2 k}} \left( 1 - \frac{i}{k \tau} \right) \;, \nonumber\\
Q_\lambda^{(0)} \left( \tau ,\, {\bm k} \right) &= \int^\tau d \tau' G_k \left( \tau ,\, \tau' \right) {\cal S}_\lambda \left( \tau ,\, {\bm k} \right) \;. 
\end{align}
The tensor power spectrum is defined as
\begin{equation}
{\cal P}_\lambda \left( k \right) \delta^{(3)} \left( {\bm k} + {\bm k'} \right) = \frac{k^3}{2 \pi^2} \left( \frac{2}{M_p a} \right)^2 \left\langle Q_\lambda \left( 0^- ,\, {\bm k} \right) Q_\lambda \left( 0^- ,\, {\bm k}' \right)  \right\rangle \;, 
\end{equation}
leading to the sum of the vacuum and sourced component
\begin{align}
&{\cal P}_\lambda^{(0)} \left( k \right) = \frac{k^3}{2 \pi^2} \frac{4}{M_p^2 a^2} \left\vert h_\lambda \left( 0^- ,\, k \right) \right\vert^2 = \frac{H^2}{\pi^2 M_p^2},\\
&{\cal P}_\lambda^{(1)} \left( k \right) \delta^{(3)} \left( {\bm k} + {\bm k}' \right) =
\frac{2 k^3}{\pi^2 M_p^2 a^2} \int_{-\infty}^{0^-} d \tau' G_k \left( 0^-,\, \tau' \right) \int_{-\infty}^{0^-} d \tau'' G_{k'} \left( 0^-,\, \tau'' \right) \left\langle {\cal S}_\lambda \left( \tau' ,\, {\bm k} \right) {\cal S}_\lambda \left( \tau'' ,\, {\bm k} \right) \right\rangle.\label{eq:tensor_sourced}
\end{align}
Similarly to the scalar power spectrum, we express the sourced component as
\begin{equation}
{\cal P}_\lambda^{(1)} \left( k \right) \equiv \frac{2}{\pi^2} \frac{H^4}{M_p^4} {\rm e}^{4 \pi \xi} f_{h,\lambda} \left( \xi \right),
\end{equation}
and by performing the integral in Eq.~\eqref{eq:tensor_sourced}, we find
\begin{equation}
f_{h,+} \left( \xi \right) \simeq  \frac{4.3 \times 10^{-7}}{\xi^6} \;\;,\;\; 
f_{h,-} \left( \xi \right) \simeq  \frac{9.2 \times 10^{-10}}{\xi^6} \;, \label{eq:fh_fit}
\end{equation}
in agreement with Eq.~(3.41) of Ref.~\cite{Barnaby:2011vw}. 

\section{Comparison of the different contributions to the scalar modes}\label{app:comparison}
\begin{figure}[ht!]
\centerline{
\includegraphics[width=0.48\textwidth,angle=0]{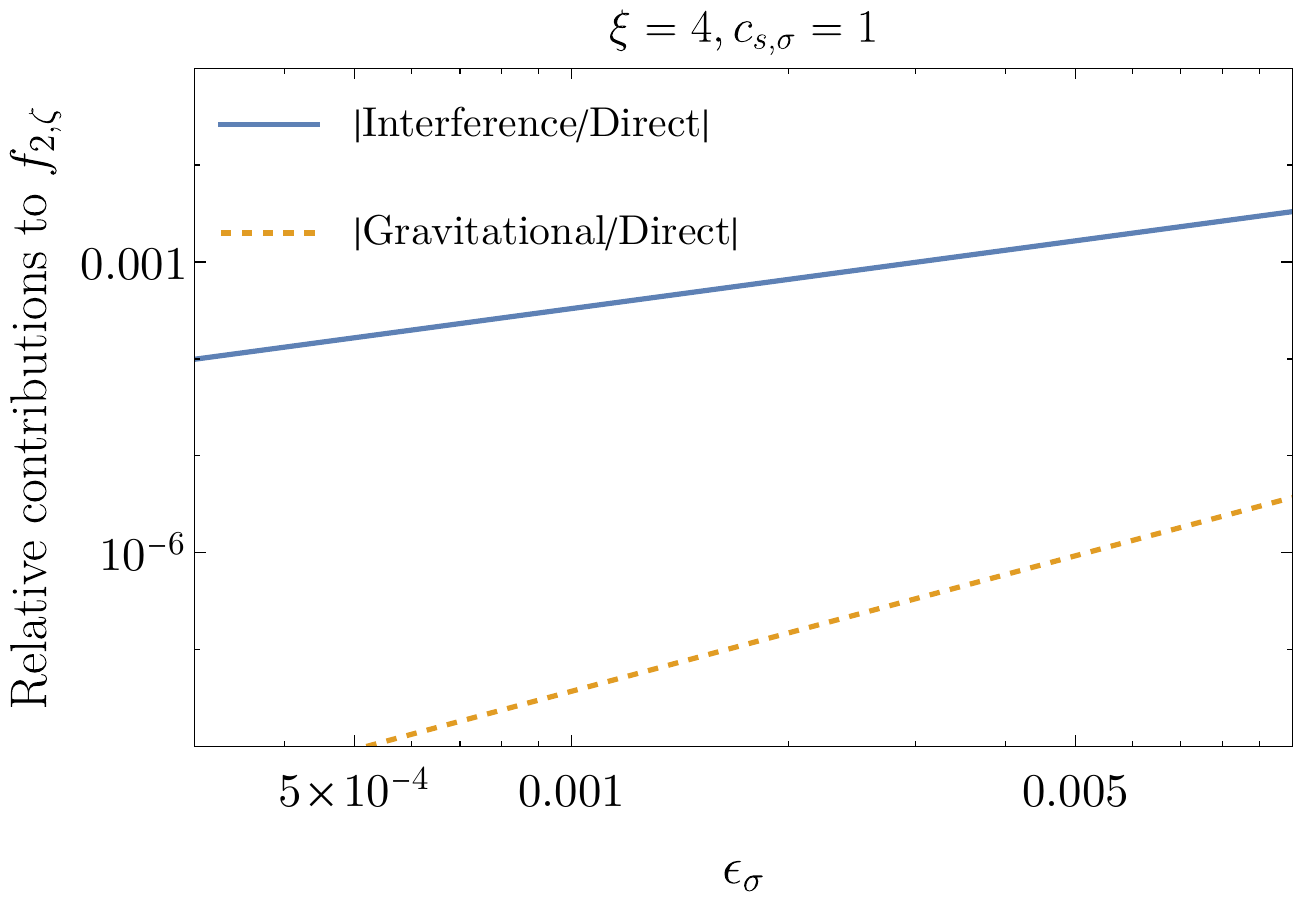}
\includegraphics[width=0.48\textwidth,angle=0]{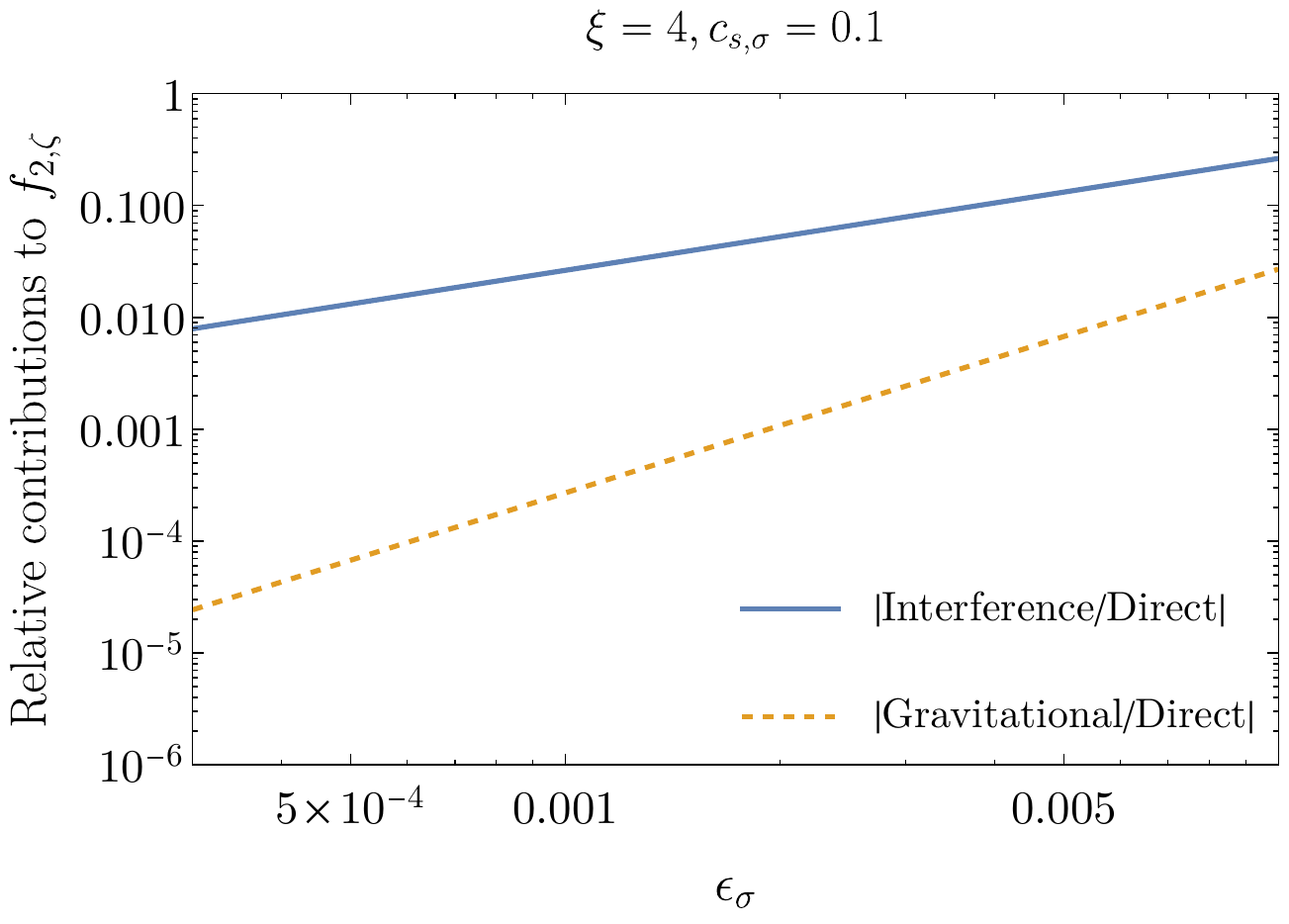}
}
\caption{The contribution of gravitational coupling and that of interference to $f_{2,\zeta}$, relative to that of direct coupling, in the single field model. 
The left and right panel correspond to $c_{s,\sigma} = 1 ,\, 0.1$, respectively, and we set the same highest value of $\epsilon_\sigma$ as the one considered in Fig.~\ref{fig:rtot}. 
As expected from Eq.~\eqref{eq:f2_full}, one can see the parametric dependence as $\epsilon_{\sigma}c_{s,\sigma}^{-2}$ and $\epsilon_{\sigma}^2 c_{s,\sigma}^{-4}$ at the leading order, respectively for the interference and the gravitational one. 
}
\label{fig:f2}
\end{figure}

Here we compare the contributions to the sourced scalar modes from the direct Chern-Simons and the gravitational coupling. In contrast to the standard scenario, the latter can have a non-negligible contribution for small sound speed.
Firstly, let us consider the single field scenario. From the fitting equation of $f_{2,\zeta}$~\eqref{eq:f2_full}, we collect the terms that depend differently on $\left\{ \epsilon_\sigma ,\, c_{s,\sigma} ,\, \xi \right\}$. 
The terms of ${\rm O } \left( \epsilon_\sigma^0 \right)$ are the direct contribution, originating from $\vec{E} \cdot \vec{B}$.  The terms of ${\rm O } \left( \epsilon_\sigma^2 \right)$ are the gravitational contribution, originating from $E^2$ and $B^2$. The terms of ${\rm O } \left( \epsilon_\sigma \right)$ are the interference between these two. We show the relative contributions of these term to $f_{2,\zeta}$ in Fig.~\ref{fig:f2}. 
For each panel, we set $\xi = 4$ 
(since the leading terms of the direct and gravitational contributions, as well as those of the interference, all scale as $\xi^{-6}$, we do not observe a significant dependence on $\xi$ of the resltive size of the various contributions) and we assume the highest value of $\epsilon_{\sigma}$ to be $10^{-2}$ as in the middle and the right panels of Fig.~\ref{fig:rtot}. 
As expected from Eq.~\eqref{eq:f2_full}, we find that the ratio between the interference and the direct contribution (resp. between the gravitational and the direct contribution) scale as $\epsilon_{\sigma}c_{s,\sigma}^{-2}$ (respectively, as $\epsilon_{\sigma}^2c_{s,\sigma}^{-4}$). This means that no contribution can be neglected at relatively large $\epsilon_\sigma$ and relatively small $c_{s,\sigma}$, as illustrated in the right panel of Fig.~\ref{fig:f2}. This can also be seen from the green lines in Fig.~\ref{fig:rtot} corresponding to the non-Gaussianity bound. While the line in the left panel (for $c_{s,\sigma} = 1$) is independent of $\epsilon_\sigma$, in the other two panels (corresponding to a smaller sound speed) the line exhibits a dependence on $\epsilon_\sigma$ at large values of this parameter, due to the contributions from the gravitational interactions. 

Finally, let us briefly discuss the two-field case.
As discussed in the main text, all contributions to the sourced scalar curvature depend on as an overall multiplicative factor at leading order (see Eqs.~\eqref{eq:source_const_xi_two_fields} and~\eqref{eq:f3_def_two_fields}). Therefore, relative weight of different contributions in this case is controlled by $\Delta N_k$ and $c_s$ as in Eqs.~\eqref{eq:f2_full_multi} and~\eqref{eq:f3_full_multi}, and there is no relative suppression by $\epsilon_{\phi}$.
By closely looking at these equations, we find that to leading order in slow-roll, the contribution of the interference relative to the direct contribution (resp. the gravitational one relative to the direct contribution) scale as $c_s^{-2}(1+c_s^2)^{-1}\Delta N_k^{-1}$ (respectively, as $c_s^{-4}(1+c_s^2)^{-2}\Delta N_k^{-2}$).
While we do not produce a plot for this two-field case, we find that, {\it e.g.}, when $\Delta N_k \sim {\cal O}(1)$ and $c_s = 0.1$, the interference and gravitational contribution becomes larger than the direct one, by a factor of more than 10. 
One can actually see this by extrapolating Fig.~\ref{fig:f2} up to $\epsilon_{\sigma} = 1$, which corresponds to the two field scenario with $\Delta N_k \sim {\cal O}(1)$.
Therefore, the contributions neglected in the standard scenario (namely, the interference and the gravitational one) can be sizable for the non-canonical scalars.
We remark that our result summarized in Fig.~\ref{fig:rtot_two_fields} was produced by taking into account all these contributions.

\bibliographystyle{JHEP}
\bibliography{ref}

\end{document}